\documentclass[a4paper,11pt]{article}
\usepackage{jcappub}
\usepackage{lineno}
\pdfoutput=1

\usepackage{xcolor}
\usepackage{upgreek}
\usepackage{subcaption}
\usepackage{dsfont}
\usepackage{hyperref}
\usepackage[normalem]{ulem}
\title{The first data-driven bounds on the quantum decoherence of inflationary gravitational waves}

\author[a,b]{Jessie de Kruijf}
\author[c,d]{Giacomo Galloni}
\author[a,b,e]{Nicola Bartolo}
\affiliation[a]{Dipartimento di Fisica Galileo Galilei, Università di Padova, I-35131 Padova, Italy}
\affiliation[b]{INFN Sezione di Padova, I-35131 Padova, Italy}
\affiliation[c]{Dipartimento di Fisica e Scienze della Terra, Università degli Studi di Ferrara, via Saragat 1, I-44122 Ferrara, Italy}
\affiliation[d]{Istituto Nazionale di Fisica Nucleare, Sezione di Ferrara, via Saragat 1, I-44122 Ferrara, Italy}
\affiliation[e]{INAF-Osservatorio Astronomico di Padova, Italy}

\emailAdd{jessiearnoldus.dekruijf@phd.unipd.it}

\abstract{
The (large-scale) structures we observe in the Universe are classical, but within the inflationary scenario they do originate from quantum fluctuations. This leads to the question: ``How did this quantum-to-classical transition occur?''. A potential explanation is quantum decoherence due to interactions between different fields present during inflation. The tensor modes (i.e. primordial gravitational waves) can interact with a  scalar sector, causing their quantum decoherence to occur and inducing a change in the gravitational wave (GW) background. The power spectrum of these GWs can be constrained using the upper bounds found by \textit{Planck}, BICEP/Keck Array, LIGO-Virgo-KAGRA, Big Bang Nucleosynthesis, and the Pulsar Timing Array detections. These impose constraints on the interaction between the fields. We find that the observational upper bounds mainly constrain scenarios with a strong interaction, especially if the interaction is also strongly time dependent. Furthermore, we find which observationally allowed scenarios have not completed decoherence by the end of inflation, thus possibly leaving quantum signatures in the GW background. Lastly, we show that, interestingly enough, there are decoherence scenarios corresponding to the signal observed by PTA experiments. This highlights the importance of the quantum decoherence effect on GWs. 
}

\begin{document}
\maketitle
\flushbottom

\section{Introduction} \label{sec: Introduction}

The theory of inflation provides a mechanism to explain the large-scale structure of the Universe, by generating primordial density (i.e. scalar ) perturbations from quantum fluctuations that were stretched to cosmological scales during a period of accelerated expansion~\cite{STAROBINSKY198099, Guth:1980zm, Mukhanov:1982nu, Bardeen:1983qw}. Crucially, inflation also predicts the generation of a stochastic background of gravitational waves (GWs) from tensor perturbations through a similar mechanism~\cite{Starobinsky:1979ty, Allen:1987bk, 1975JETP...40..409G}, namely, the amplification of quantum vacuum fluctuations. 

The general expectation is that these density perturbations could have originated from quantum fluctuations if these fluctuations were classicalized in some way, meaning that a quantum-to-classical transition must have occurred in a cosmological context~\cite{Perez:2005gh,Ashtekar:2020, Guth:1985ya, Polarski:1995jg}.
How this transition happens exactly is not clear, although the problem has been studied extensively, see e.g.~\cite{ Kiefer:2008ku,Lesgourgues:1996jc, Kiefer:1998qe, Sudarsky:2009za,Das:2013qwa,Chandran_2024}. A possible explanation is the quantum decoherence of the perturbations~\cite{Brune:1996zz,Schlosshauer:2019ewh,Burgess_2008,Schlosshauer:2014pgr,Barvinsky:1998cq, Zurek:1981xq, Joos1985TheEO, Schlosshauer_2005,10.1093/acprof:oso/9780199213900.001.0001, Alicki_2004,cespedes2025cosmologydecoherencesecondlaw, sano2025falsegenuinedecoherenceearly, lopez2025quantumsignaturesdecoherenceinflation,Burgess_2023, 2025arXiv250907769B, 2024arXiv240312240B, 2024JCAP...05..025C}, where the perturbations are treated as an open quantum system, interacting with an environment (e.g. other fields present in the early universe). This interaction suppresses the off-diagonal elements of the density matrix and drives the system toward classicality \cite{Burgess_2008, 1801.09949}. 

In our previous work~\cite{de_Kruijf_2024}, we studied how quantum decoherence affects the primordial gravitational wave background (GWB). Specifically, we used the formalism introduced in~\cite{1801.09949} to study the interaction between inflationary tensor perturbations and an unspecified scalar environment present during inflation. We obtained an analytical expression for the change to the primordial GWB power spectrum that quantum decoherence induces. Additionally, we used upper bounds on the tensor-to-scalar ratio and tensor spectral index, found by~\cite{Galloni_2023}, to find approximate constraints on some of the quantum decoherence parameters. Finally, we investigated on which scales decoherence might not have completed before the end of inflation, since this leaves the interesting possibility that quantum signatures might have survived and left imprinted in the GWB.  

We will extend on that work here by performing a full parameter space analysis, using multi-scale observational data on the inflationary GWs. On CMB scales, the tightest bounds are provided by data from the BICEP/Keck Array and the \textit{Planck} satellite~\cite{ 2018PhRvL.121v1301B, 2020A&A...641A...6P, 2021A&A...647A.128T, 2021PhRvL.127o1301A, 2022PhRvD.105h3524T, Galloni_2023, 2014A&A...571A..16P, Planck:2015fie,Planck:2018jri, Ade_2021, Galloni_2023, Galloni2024, Tristram_2022}. 
On much smaller scales, the upper bound is set by the direct GW observations of  ground-based interferometers, Laser Interferometer Gravitational-Wave Observatory (LIGO), Virgo, and Kamioka Gravitational Wave Detector (KAGRA), hereafter referred to as LVK~\cite{PhysRevLett.118.121101, PhysRevD.104.022005, Abbott_2021, theligoscientificcollaboration2025upperlimitsisotropicgravitationalwave}. Due to the possible scale-dependence of the decoherence effect on the primordial tensor power spectrum~\cite{de_Kruijf_2024}, the combination of these data sets can be used to constrain the quantum decoherence parameters. Additionally, Big Bang Nucleosynthesis (BBN) provides an upper limit on the smallest scales through its constraining power on the effective number of relativistic particle species.
Finally,  the Pulsar Timing Array (PTA) method~\cite{Verbiest_2021, 2023ApJ...951L...8A, 2023arXiv230616226A, 2023ApJ...951L...6R, InternationalPulsarTimingArray:2023mzf, Xu_2023}, has found evidence of a stochastic GWB, on intermediate scales. The origin of the signal has not been definitively proven; thus, we will perform two types of analysis. First, we use the found signal as an upper bound on the inflationary GWs, assuming that the signal has a different origin. Secondly, we will look at which decoherence scenarios could explain the PTA signal. 

Furthermore, we combine the observational constraints with theoretical assumptions from the decoherence criterion. The criterion indicates how mixed the state of the system has become due to the interaction with the environment~\cite{Burgess_2008}, depending on the interaction strength between the system and the environment. Assuming the largest scales have been fully decohered gives a lower bound on the interaction strength. However, smaller scales decohere for a shorter amount of time, due to them exiting the environmental correlation length at a later time. If by the end of inflation the decoherence is incomplete, these scales could still show some quantum signatures. Studying these very small scales, like LVK scales, allows us to probe which decoherence scenarios could still retain some quantum signatures~\cite{de_Kruijf_2024}. A broad range of quantum signatures could still be present, with several works focusing on how to detect them (see e.g.~\cite{2021PhRvD.103d4017K, micheli2023quantum, 2023arXiv230202584N, Campo:2005sv,Martin2023, PhysRevD.61.024024, Banerjee_2023, Sharifian_2024, Parikh_2020, Parikh_2021, Parikh_2021_signatures, Lamine_2006, Choudhury_2017}).

This work is organized as follows. We start in Section~\ref{sec: quantum decoherence} with a brief overview of quantum decoherence and the analytical results found in~\cite{de_Kruijf_2024}. In Section~\ref{sec: MCMC} we give details on the Markov-Chain Monte-Carlo (MCMC) framework used to perform the analysis. Then, section~\ref{sec: data sets} expands on the data sets used for the analysis presented in this work, and section~\ref{sec: results} shows the results of this analysis. Lastly, in Section~\ref{sec: Conclusion}, we summarize these results.  


\section{Quantum decoherence}\label{sec: quantum decoherence}

Quantum decoherence is the process with which a quantum system loses its coherence, due to interactions with an environment. One way to describe this process is to derive a Lindblad equation~\cite{Burgess_2008, 1801.09949, Shandera:2017qkg, Pearle_2012, Brasil_2013}, under some typical assumptions and approximations \footnote{The derivation assumes the quantum decoherence process is Markovian and that the process has no feedback on the environment.}. This equation describes the time evolution of the (reduced) density matrix of a system that interacts with an environment. The interaction suppresses the off-diagonal elements of the density matrix of the system, but also affects the diagonal elements.
This means that the statistical properties of the system, i.e. the observable predictions for its power spectra and higher-order correlation functions, can be modified~\cite{Burgess_2008, Boyanovsky_2015,  1801.09949}.
In our scenario, the system consists of the tensorial modes present during inflation, and the environment is a generic scalar sector. Indeed, in the derivation of the Lindblad equation, the environment is traced out, which allows us, within an effective field theory approach, to not specify it, but rather model it under some generic premises and considerations (for more details on this point, see ~\cite{1801.09949,de_Kruijf_2024}). 

\subsection{Tensor Power spectrum}\label{sec: tensor power spectrum}

The full derivation of the GW power spectrum including decoherence effects can be found in~\cite{de_Kruijf_2024}; here, we only summarize the main results and parameters. 
The GW power spectrum, including the effect due to quantum decoherence, at a given scale $k$, is
\begin{equation}\label{eq: PT full}
\begin{aligned}
    \mathcal{P}_{T}(k) = \mathcal{P}_{T|\mathrm{stan}} \Bigg( 1 + \frac{2 \beta^2 \sigma_{\gamma}}{9 \sin^2(\nu\pi)} \bigg( \frac{k}{k_{\ast}} \bigg)^{p+1} \frac{JI(k, \eta, \nu)}{[J_{\nu}^2(-k\eta) + Y_{\nu}^2(-k\eta)]} \Bigg),
\end{aligned}
\end{equation}
with $J_{\nu}(z)$ and $Y_\nu(z)$ the Bessel function of the first and second kind respectively, where $\nu$ is equal to $3/2$ when neglecting slow-roll parameters. 
Here 
\begin{equation}\label{eq: standard PS}
    \mathcal{P}_{T|\mathrm{standard}} = A_s r \left( \frac{k}{k_\ast} \right)^{n_T}
\end{equation} is the ``standard'' GW power spectrum, without quantum decoherence, where $A_s$ is the amplitude of the scalar power spectrum at the pivot scale $k_\ast =0.05~~\mathrm{Mpc}^{-1}$, $r$ is the tensor-to-scalar ratio, and $n_T$ is the tensor spectral tilt.

In this context, the power spectrum depends on several quantum decoherence parameters. First, $p$, which is used to parameterize the time dependence of the interaction between the system and the environment. 
To be exact, the interaction term of the Lindblad equation is dependent on the coupling constant $g$ and the autocorrelation (conformal) time of the environment $\eta_c$, which is parameterized as\footnote{The time-dependent coefficient of the interaction term in the Lindblad equation is given by $\gamma=g^2\eta_c$, as can be seen in~\cite{1801.09949, de_Kruijf_2024}.}
\begin{equation}\label{eq: gamma dependence on a}
   \gamma = \gamma_{\ast} \Bigl( \frac{a}{a_{\ast}} \Bigr)^p,
\end{equation}
where $a$ is the scale factor and $\ast$ refers to a reference time which is taken to be the time when the pivot scale $k_{\ast} = 0.05~\mathrm{Mpc}^{-1}$ crosses the Hubble radius during inflation (i.e. $k_{\ast} = a_{\ast}H_{\ast}$).

Secondly, the power spectrum depends on the strength of the interaction with the environment, which is parameterized as $\beta^2\sigma_{\gamma}$. Here, $\sigma_{\gamma}$ is defined as
\begin{equation}
    \sigma_{\gamma} \equiv  \frac{l_E^3\Bar{C}_R\gamma_{\ast}k_{\ast}^{4}}{a_{\ast}^3},
\end{equation}
with $l_E$ the correlation length of the environment, $\Bar{C}_R$ the amplitude of the spatial correlation of the environment and $\beta = \frac{2\xi}{M_{pl}^4}$ is a dimensionless constant, with $\xi^2$ used as an expansion parameter and $M_{pl}$ the Planck mass (used to maintain the analogy with \cite{1801.09949, 2211.07598}). 
Finally, the product $JI(k, \eta, \nu)$ of Eq.~\ref{eq: PT full} is defined as
\begin{equation}
    JI(k, \eta, \nu) =  J_{-\nu}^2(-k\eta)I_1(\nu, k, \eta) -2J_{-\nu}(-k\eta)J_{\nu}(-k\eta)I_2(\nu, k, \eta) + J_{\nu}^2(-k\eta)I_1(-\nu, k, \eta) ,
\end{equation}
with the integrals $I_1$ and $I_2$ defined by
\begin{equation}\label{eq:I1,I2 definition}
\begin{aligned}
    &I_1(\nu) \equiv \int_{-k\eta}^{(H_{\ast}l_E)^{-1}} \mathrm{d}z z^{2-p} \bigg( \frac{1}{z^2} - \frac{1}{(-k\eta_{IR})^2} \bigg) J_{\nu}^2(z), \\
    &I_2(\nu) \equiv \int_{-k\eta}^{(H_{\ast}l_E)^{-1}} \mathrm{d}z z^{2-p} \bigg( \frac{1}{z^2} - \frac{1}{(-k\eta_{IR})^2} \bigg) J_{\nu}(z)J_{-\nu}(z).
\end{aligned}
\end{equation}
Furthermore, $\eta_{IR}$ is defined through the number of $e$-folds elapsed since the onset of inflation $N-N_{IR} \equiv \ln{(\eta_{IR}/\eta)}$, and parameterized as the amount of $e$-folds from the start of inflation until the end $N_T = N_{end} - N_{IR}$. For this analysis, we set $N_T = 10^4$ (by definition this number needs to be larger than number of $e$-folds since the pivot scale crossed the horizon $N_\ast$). From Eq. (\ref{eq:I1,I2 definition}) it is clear that for those values, the dependence on $N_T$ is negligible. Therefore do not vary it in our analysis.

The upper limit of integration in Eq.~\ref{eq:I1,I2 definition} corresponds to the time when the wavelength ($a/k$) of the co-moving mode under consideration $k$ crosses the correlation length of the environment $l_E$, at leading order, this is $-k\eta_E = (H_{\ast}l_E)^{-1}$. This means the effect of quantum decoherence is only present on scales large enough to feel the effect of the environment. Therefore $l_E$ provides a natural cut-off for the modifications to the standard power-spectrum. Let us note that in the derivation of the Lindblad equation, it is assumed that $H_{\ast}l_E \ll 1$ and this will be enforced throughout this work. Then, defining $ \Delta N_{\ast} = N_{end}-N_{\ast}$ as the number of $e$-folds elapsed since the pivot scale crossed out of the Hubble radius until the end of inflation, we can write the scales which are influenced by quantum decoherence as
\begin{equation}\label{eq: dec cut-off}
    \frac{k}{k_{\ast}} < (H_\ast l_E)^{-1} e^{\Delta N_{\ast}}.
\end{equation}
For the scenarios considered in~\cite{de_Kruijf_2024}, the cut-off scale of the decoherence contribution was on scales much smaller than the ones probed by LVK. However, in this work, we want to expand the analysis to also include more diverse models of inflation (varying $\Delta N_{\ast}$), and probing different environmental correlation lengths (varying $ H_\ast l_E$).  
When varying $\Delta N_{\ast}$, the amount of observable modes crossing out of the horizon during inflation is affected. In particular the smallest mode to have crossed out of the horizon is\footnote{This expression neglects any effects from reheating or effects after inflation. A more detailed expression is given in~\cite{Liddle_2003} but goes beyond the scope of this work. However we do take them effectively into account by the choice of our prior range for $\Delta N_{\ast}$. }
\begin{equation}\label{eq: k max}
    k_{\rm max} \simeq H_0 e^{N_{\rm obs}},
\end{equation}
where $H_0$ is the present Hubble constant, and $k_{\rm max}$ depends on the the minimum number of $e$-folds (for successful inflation), corresponding to the largest observational scales 
\begin{equation}\label{eq:N_obs}
    N_{\rm obs} = \Delta N_\ast + \ln{\left( \frac{k_\ast}{H_0} \right)} \simeq \Delta N_\ast + 5.4.
\end{equation}
Typically, $k_{\rm max}$ is considered as a limit 
beyond which it is not possible to excite fluctuations modes by the accelerated expansion taking place during inflation. 
Notice however, that in principle the correlation length of the environment $l_{\rm E}$  can be smaller than the size of the horizon at the end of inflation determined by~(\ref{eq: k max}). Indeed, since we neglect the slow-roll parameters and assume $H_\ast l_E \ll 1$, this is the case. 
It would therefore make sense to address the possible contributions from modes $k> k_{\rm max}$ as well, until the cut-off $l_{\rm E}$ takes over.
However, since this concerns the ultraviolet regime, one needs to properly take it into account when computing the quantum decoherence contribution, and the accurate way of doing so is a heavily debated subject.\footnote{See the discussions in, e.g.,~\cite{1801.09949,lopez2025quantumsignaturesdecoherenceinflation} and Refs. therein.}  We do therefore prefer to impose a second cut-off for scales $k> k_{\rm max}$, adopting however such a cut-off as a proxy for the one that would be naturally played in any case by $l_{\rm E}$ (the two indeed are equal in the case when $a_{\rm end}/k_{\rm max}=l_{\rm E}$). The natural cut-off provided by the environment could in fact play a crucial role when we try to fit the PTA data within such scenarios, still maintaining observational consistency with all the various constraints (CMB, LVK and BBN bounds). An in-depth study of the possible ultra-violet contributions is left for future work.

\subsection{Decoherence criterion}

The loss of coherence, and the purity of the state~\cite{Burgess_2008,1801.09949, 2024arXiv240312240B}, can be described in terms of the quantity $\delta_{\boldsymbol{k}}$ which characterizes the additional decrease in off-diagonal elements of the reduced density matrix of the system, due to the interaction with the environment.
In our case, this parameter is given by~\cite{de_Kruijf_2024}
\begin{equation}\label{eq: decoherence criterion}
    \delta_{\boldsymbol{k}}(\eta) = \frac{2\beta^2\sigma_{\gamma}}{9\sin^{2}{(\nu\pi)}} \bigg( \frac{k}{k_{\ast}} \bigg)^{p+1} \big[I_{1}(\nu, k, \eta) + I_{1}(-\nu, k, \eta) - 2I_{2}(\nu, k, \eta)\cos{(\nu\pi)}\big].
\end{equation}
Successful decoherence is characterized by the condition $\delta_{\boldsymbol{k}} \gg 1$, referred to as the decoherence criterion. 
As can be seen from its definition, this criterion is scale-dependent. The smallest scales spend the shortest amount of time outside of the environmental correlation length, meaning they only decohere for a short amount of time, and it is not guaranteed that they will fully decohere before the end of inflation. The largest scales, however, spend much more time decohering, and it is reasonable to assume that by the end of inflation, they have classicalized. Hence, we can impose that the cosmic microwave background (CMB) scales have fully decohered ($\delta_{\boldsymbol{k}_{CMB}} \gg 1$).
Nevertheless, let us note that the cosmological GWB has not been detected yet, meaning we can not be sure that these modes have indeed become classical. Aside from that, imposing the decoherence criterion on CMB scales means the tensor modes must have classicalized only due to the interactions with a scalar environment (as is detailed in~\cite{de_Kruijf_2024}), not by any other means.
Since we do not know how many fields were present in the early Universe, and there are potential other ways to fully decohere inflationary perturbations, e.g. \cite{2025arXiv250907769B, 2025arXiv250907077C, takeda2025quantumdecoherencegravitationalwaves, Nelson_2016, 2025JCAP...05..032S, lopez2025quantumsignaturesdecoherenceinflation, Schlosshauer:2014pgr, Burgess_2023} 
this could be a simplification. Because of these caveats, we show our results both with (Section~\ref{sec: results decoherence criterion}) and without (Section~\ref{sec: results r, p, sigma}) assuming the decoherence criterion. 
Finally, another reason to consider this criterion is to determine whether there are scales on which tensor modes have not fully decohered~\cite{de_Kruijf_2024} (see Section~\ref{sec: results decoherence criterion}). This would open up a window into testing the possible quantum signatures in the GW background~\cite{2021PhRvD.103d4017K, micheli2023quantum, 2023arXiv230202584N, Campo:2005sv, Martin2023, PhysRevD.61.024024, Banerjee_2023, Sharifian_2024, Parikh_2020, Parikh_2021, Parikh_2021_signatures, Lamine_2006, Choudhury_2017}.~\footnote{However, see
\cite{Martin2023, Martin:2022kph}, where, even in the case where decoherence has been reached, a criterion is provided to understand when it is still possible to extract quantum signatures from the system.}


\section{Methodology and parameter priors}\label{sec: MCMC}

We use a Monte Carlo Markov Chain (MCMC) analysis~\cite{Metropolis_1949, Lewis_2002} to extract information on the cosmological parameters from the data. In particular, chains are run using \texttt{cobaya}~\cite{Torrado:2020dgo}\footnote{\href{https://github.com/CobayaSampler/cobaya}{https://github.com/CobayaSampler/cobaya}}, and we utilize the cosmological theory code \texttt{CAMB}~\cite{Lewis:1999bs, Lewis:2002ah, Howlett:2012mh}\footnote{\href{https://github.com/cmbant/CAMB}{https://github.com/cmbant/CAMB}} and the chains are studied using \texttt{GetDist}~\cite{Lewis:2019xzd}.

This work focuses entirely on the effect of quantum decoherence of tensorial modes during inflation, due to a scalar (field) environment present during inflation. Therefore, we do not vary the 6 $\Lambda$CDM parameters ($A_s$, $n_s$, $\Omega_b h^2$, $\Omega_{cdm}h^2$, $\theta_s$, $\tau_{\rm reio}$) and set them to the values found by \textit{Planck}~\cite{Planck:2018vyg}.\footnote{We have confirmed that opening up the parameter space, to include the $\Lambda$CDM parameters, does not significantly affect our results, as can be seen in Appendix~\ref{sec: LCDM}.}
An overview of the parameters studied in this work, and the priors imposed on them, is shown in Table~\ref{tab: priors parameters}. 

\begin{table}
    \centering
    \begin{tabular}{|c|c|}
        \hline
        Parameters & Prior range \\
        \hline
        $r_{0.05}$ & [0,3] \\
        $n_T$ & [-5,5] \\
        \hline
        $p$ & [-10,10] \\
        $\log_{10}(\beta^2\sigma_\gamma) $ & [-40,20]\\
        $\log_{10}(H_\ast l_E)$ & [-30,-1] \\
        $\Delta N_\ast$ & [12.6, 101.6] \\
        \hline
    \end{tabular}
    \caption{The prior range for the decoherence and tensorial parameters, unless stated otherwise. }
    \label{tab: priors parameters}
\end{table}

\subsection{Standard power spectrum}

The inflationary tensor power spectrum, without decoherence, is given by Eq.~\eqref{eq: standard PS}.
In this work, we assume uniform priors, $0 < r_{0.05} < 3$, $-5 < n_T < 5$. 
Of course, $n_T$ becomes more and more unconstrained as $r_{0.05}$ tends to zero, causing the posterior distribution to suffer prior-volume effects. These were studied in detail in~\cite{Galloni_2023, Galloni2024}. Thus, to limit them, we adopt the same approach to the tensor sector sampling as in those references.

Even though the cosmological GWB has not been detected yet, one often assumes the value predicted by the Starobinsky model as a reference value, namely $r_{0.05} = 0.00461$~\cite{STAROBINSKY198099, LiteBIRD:2022cnt}. Given that this prediction sits in the middle of the $r-n_S$ posterior extracted from current CMB data \cite{Galloni_2023, Planck:2018jri}, it is a natural target for future CMB missions. Thus, in Section~\ref{sec: results Starobinsky}, we will assume such a value for $r_{0.05}$ to gain some intuition on the consequences of Starobinsky inflation on the decoherence parameters.

\subsection{Decoherence parameters}

The priors on the decoherence parameters, summarized in Table \ref{tab: priors parameters},
are all uniform and chosen based on analytical and numerical results found in~\cite{de_Kruijf_2024}. First, the time dependence of the interaction strength $p$ (see Eq.~\ref{eq: gamma dependence on a}) could, in theory, span over a large range of values, but we do not expect the interaction to very strongly depend on time. Specifically, $p=-1$ corresponds to a quantum decoherence scenario that produces a scale-invariant contribution to the tensor-power spectrum. It can originate from an interaction between the tensor and scalar inflationary perturbations (two gravitons and the curvature perturbation) as described by the cubic term of the single field slow-roll action, given in~\cite{Maldacena2003}. Aside from that, the value of $p+1$ determines the slope of the quantum decoherence induced contribution to the tensor power spectrum\footnote{As can be seen from Eq.~\eqref{eq: PT full}, where the $p$ dependence coming from the integrals given in Eq.~\eqref{eq:I1,I2 definition} is small.}, meaning that too extreme values will disagree with the current data. Therefore, to limit the prior range, while being able to observe this behavior, we have chosen to probe the values $-10 < p < 10$.

Secondly, treating the interaction strength $\beta^2\sigma_\gamma$ as a phenomenological parameter, we do not impose any theoretical upper bound on it; however,~\cite{de_Kruijf_2024} gives us an intuition on the expected observational bound. Thus, to avoid a too stringent prior, we assume $\beta^2\sigma_\gamma \leq 10^{20}$.
On the other hand, the decoherence criterion can provide a lower bound. As mentioned above, there are caveats to this lower bound, so we study how the observational constraints change with or without this criterion. We impose a lower bound of  $\beta^2\sigma_\gamma \geq 10^{-40}$, since we do not expect any extra behavior beyond what is characterized in this range.

As mentioned above, the parameter $H_\ast l_E$ describes the characteristic size of the correlation length of the environment (w.r.t. the Hubble radius), but since the environment is traced out, this is a free parameter. However, in the derivation of the Lindblad equation~\cite{1801.09949, de_Kruijf_2024}, $H_\ast l_E \ll 1$ is assumed, given that such a correlation length must be smaller than  the Hubble radius.
Practically, this means we set $H_\ast l_E = 0.1$ as an upper bound, to ensure that the assumption holds. There is no theoretical lower bound, but to avoid strong volume effects, we set the lower limit to  $H_\ast l_E = 10^{-30}$. Note that, to our knowledge, this is the first time observational bounds are used to study the correlation length of the environment using data on stochastic GW backgrounds. This could potentially provide specific insights into the physical properties of the environment.

Lastly, for the amount of $e$-folds between the pivot scale exiting the horizon and the end of inflation ($\Delta N_\ast$), we use the theoretical constraints on $N_{\rm obs}$ found in~\cite{Liddle_2003}, converted to our parameter using Eq.\eqref{eq:N_obs}. In principle, these constraints are dependent on the energy scale of inflation and the details of the reheating phase. 
A more in-depth analysis on the connection between quantum decoherence and reheating can be found in~\cite{takeda2025quantumdecoherencegravitationalwaves}, but this is beyond the scope of this work. The prior we enforce is $ 12.6 \lesssim \Delta N_{\ast} \lesssim 101.6$. As discussed in~\cite{Liddle_2003}, the lower and upper limits in general correspond to extreme models of inflation; thus, this range completely encodes the variability of $\Delta N_{\ast}$.  


\section{Datasets}\label{sec: data sets}

Let us give a quick overview of the data used to obtain the bounds on the primordial GW power spectrum, divided according to different detectors.

\subsection{\textit{Planck} satellite}

The \textit{Planck} satellite was designed principally to deliver unprecedented accuracy in measurements of the CMB temperature anisotropies, while also advancing measurements of CMB polarization.
As of today, it has provided the tightest constraints on the $\Lambda$CDM parameters. Over recent years, this legacy has been extended by ground-based observatories. For example, the recent data release of the Atacama Cosmology Telescope (ACT) ~\cite{Naess_2025, ACT:2025fju} and the South Pole Telescope (SPT) \cite{SPT-3G:2024atg, SPT-3G:2025bzu}. Combining \textit{Planck} with these experiments allows one to surpass the capabilities of \textit{Planck} alone in tightening cosmological constraints. However, these mainly affect the constraints on the scalar part of the parameter space, only marginally considered here, and therefore are not relevant to this work.

We use the data from both the Planck Release 3 (PR3)~\cite{2020_Planck_V} and the Planck Release 4 (PR4), which uses the NPIPE end-to-end pipeline~\cite{Rosenberg_2022, Tristram_2024}. Let us note that the two releases (PR3 and PR4) are not independent. Thus, we select our likelihoods in order to avoid double-counting of multipole information. In particular, we isolate the PR4 contribution to the low-$\ell$ part of the polarization fields: the ``Low-L Likelihood Polarized for \textit{Planck}'' (LoLLiPoP) likelihood\footnote{https://github.com/planck-npipe/lollipop}, describing the low-$\ell$ spectrum of EE, EB, and BB \cite{Couchot_2017, Couchot_2017_cosmo, Mangilli_2015}. We combine it with the low-$\ell$ part of the CMB temperature, named ``lowlTT'', describing the low-$\ell$ parts of the TT spectrum~\cite{2020_Planck_V}. Finally, we include the PR4 CMB lensing likelihood~\cite{Carron_2022}. To simplify the notation, we will refer to these likelihoods together as ``PL''. Other PR3 and PR4 likelihoods are not used in the main body of this work, since we keep the standard $\Lambda$CDM parameters fixed. This is expanded upon in Appendix~\ref{sec: LCDM}.

\subsection{BICEP/Keck array}

The BICEP/Keck array represents the present foremost attempt at measuring the primordial tensor perturbations, proving to be crucial in obtaining our current knowledge. The latest released data is publicly available through a likelihood called ``BK18''~\cite{Ade_2021}. Together with the \textit{Planck} satellite data, it has managed to constrain $r_{0.05}<0.032$ at 95\% Confidence Level (CL)~\cite{Tristram_2022}. These results were obtained by keeping $n_T$ fixed using the single-field slow-roll consistency relation ($r = -8 n_T$). When $n_T$ is left to vary, the constraints are $r_{0.01}< 0.028$ and $-1.37 < n_T < 0.42$~\cite{Galloni_2023}.\footnote{This result was obtained using the data from LVK as well, as described in the following section.}

\subsection{LIGO-Virgo-KAGRA}

Another upper bound on the stochastic GW background comes from the ground-based GW interferometers, LIGO, Virgo, and KAGRA~\cite{PhysRevLett.118.121101,  Abbott_2021}. 
These instruments have detected individual GW events but have not yet been able to detect a background. Complementarily to CMB data, the interferometers probe very small scales ($k \sim 10^{16}~\mathrm{Mpc}^{-1}$). This means that by combining both CMB and direct interferometric measurements, we can probe scales over a range of $\sim 18$ orders of magnitude. Let us note that for simplicity we take $n_T$ to be constant on this entire range of scales, but it could be allowed to vary~\cite{Planck:2018jri, Giare_2021}.

The upper bound reported after the third observing run of LVK is $\Omega_{\rm{GW}}^{\rm LVK} (25 \rm{Hz}) < 6.6 \times 10^{-9}$ at 95\% CL, obtained by marginalizing over the spectral tilt~\cite{Abbott_2021}.\footnote{While writing this paper, new results have come out for the upper bound on the (inflationary) GW background from LVK~\cite{theligoscientificcollaboration2025upperlimitsisotropicgravitationalwave} In particular, the reported upper bound shrank by a factor $2.1\times$. We tested the impact of this on our analysis and found no significant changes; thus, our results should be considered up to date.}
To encode this result into a likelihood, we use the approach shown in~\cite{Planck:2018jri, Galloni_2023}, and define a Gaussian likelihood for LVK, namely,
\begin{equation}
    -2 \log (\mathcal{L}_{\rm LVK}) = \frac{(\Omega_{\rm{GW}} - \mu_{\rm LVK})^2}{\sigma_{\rm LVK}^2}.
\end{equation}
The likelihood is centered in $\mu_{\rm LVK}=0$ and has a $\sigma_{\rm LVK}$ dispersion, which is half of the 95\% bound provided by LVK. 

The theoretical value of $\Omega_{\rm{GW}}$, as a function of other parameters, is obtained throughout our MCMC pipeline with this approximation~\cite{2018CQGra..35p3001C}
\begin{equation}
    \Omega_{\rm{GW}}(k) = \frac{3}{128} \Omega_{rad} \mathcal{P}_T(k) \left[ \frac{1}{2} \left(\frac{k_{eq}}{k} \right)^2 + \frac{16}{9} \right],
\end{equation}
where $\Omega_{rad}$ is the radiation energy density today, and $k_{eq}$ the scale entering the horizon at matter-radiation equality.

\subsection{NANOGrav}

Utilizing the Pulsar Timing Array (PTA) method~\cite{Verbiest_2021}, various PTA collaborations, namely, NANOGrav, EPTA/InPTA, PPTA, and CPTA, have found evidence of a GW background~\cite{2023ApJ...951L...8A, 2023arXiv230616226A, 2023ApJ...951L...6R, InternationalPulsarTimingArray:2023mzf,Xu_2023, EPTA_2023}.
The PTA signal probes scales around $k \sim 10^7~\mathrm{Mpc}^{-1}$, being a complementary probe on intermediate scales between CMB and LVK.
Considering that the signals from the different PTA collaborations are consistent with each other, and the fact that NANOGrav determined the spectral index with the highest precision, in this work, we only use the NANOGrav 15yr results. We expect the results we find to be consistent with the other PTA experiments, as is discussed in~\cite{Vagnozzi_2023, Figueroa:2023zhu}. 
The GWB has a total integrated energy density $\Omega_{\rm{GW}}^{\rm NANO} = 9.3^{+5.8}_{-4.0} \times 10^{-9}$ at frequency $f_{\rm yr}=1 \ \rm{yr}^{-1}$, assuming the fiducial power-law model ($f_{\rm yr}^{-2/3}$ for the characteristic strain)~\cite{2023ApJ...951L...8A}. 

Unfortunately, it has not been definitively determined whether the signal is of cosmological or astrophysical origin~\cite{Nano_new_physics,Figueroa:2023zhu, bian2023gravitationalwavesourcespulsar,Vagnozzi_2023, 2024SCPMA..6740412W, 2024PhRvD.109b3522E, EPTA_2023}. 
The standard interpretation is that the signal is mainly due to supermassive black holes~\cite{Agazie_2023}, but this has not been conclusively proven; indeed, at the moment cosmological sources for such a background are viable possibilities (see, e.g.~\cite{Figueroa:2023zhu} and~\cite{Nano_new_physics})
Therefore, we use this detection in two ways, as illustrated below.

First, we assume the cosmological contribution is subdominant, and we take it as another upper bound on the cosmological GWB. We apply the same method as was used for the LVK bounds, providing us with $\mathcal{L}_{\rm NANO}^{\rm bound}$, now with $\mu_{\rm NANO}^{\rm bound}=0$ and $\sigma_{\rm NANO}^{\rm bound}$ half of $\Omega_{\rm{GW}}^{\rm NANO}$. 

Secondly, we consider which quantum decoherence scenarios can explain the entirety of NANOGrav detection. This is done using the approach sketched out in~\cite{Galloni_2023}. The NANOGrav data is reported in terms of Amplitude ($A_{\rm GW}$) and spectral tilt ($\gamma$) which we translate into $\Omega_{\rm{GW}}$ using 
\begin{equation}
    \Omega_{\rm{GW}} (f_{\rm yr}) = \frac{2 \pi^2}{3 H_0^2} f_{\rm yr}^2 A_{\rm GW}^2,
\end{equation}
which is independent from $\gamma$. We have verified that the logarithm of $\Omega_{\rm{GW}}$ approximately follows a Gaussian distribution. Fitting the data to this distribution gives $\log (\Omega_{\rm{GW}}) = -7.3 \pm 0.4$.\footnote{The dataset and further information can be found at \href{https://github.com/nanograv/15yr_stochastic_analysis}{https://github.com/nanograv/15yr\_stochastic\_analysis} } The likelihood used by the MCMC then reads
\begin{equation}
    -2 \log (\mathcal{L}_{\rm NANO}^{\rm det.}) = \frac{(\log(\Omega_{\rm{GW}}) - \mu_{\rm NANO})^2}{\sigma_{\rm NANO}^2},
\end{equation}
with $\mu_{\rm NANO}^{\rm det.} = -7.3$ and $\sigma_{\rm NANO}^{\rm det.} = 0.4$. 

\subsection{Big Bang Nucleosynthesis}

In addition to the datasets mentioned above, BBN can also be used to constrain the gravitational wave power spectrum, especially on extremely small scales \cite{Maggiore_2000, Boyle_2008, Servant_2024}. BBN predicts the primordial abundances of Helium and other light nuclei, depending on the effective number of relativistic particle species ($\Delta N_{\rm{eff}}$). Since the total energy density contains a contribution from primordial GWs, the primordial abundance predicted by BBN can be used to constrain the primordial GW power spectrum. On large scales, this bound is much weaker than those from the CMB, LVK or the PTA method, but on very small scales, these constraints are significant. There are proposals for several future instruments which also probe these extremely small scales (see e.g.~\cite{Servant_2024, 2021LRR....24....4A, Aggarwal:2025noe}). However, as of now, none of these instruments reach the sensitivity of BBN for the primordial GW power spectrum, so we only focus on BBN constraints. 

The upper bound on the energy-density fraction in GWs today spanning in frequency $f_{\rm{GW}}$ can be written as~\cite{Servant_2024, Planck:2018jri, 2018CQGra..35p3001C, Boyle_2008, Allen_1999, Smith_2006, Kuroyanagi_2015, Ben_Dayan_2019}:
\begin{equation}\label{eq: BBN bound}
    \int_{f_{\rm{GW}}^{\rm{min}}}^{f_{\rm{GW}}^{\rm{max}}} \frac{\mathrm{d}f_{\rm{GW}}}{f_{\rm{GW}}} h^2 \Omega_{\rm{GW}} (f_{\rm{GW}}) \lesssim 5.6 \times 10^{-6} \Delta N_{\rm{eff}},
\end{equation}
where $h\simeq 0.68$ and $f_{\rm{GW}}^{\rm{min}} (f_{\rm{GW}}^{\rm{max}})$ is the lower (upper) cut-off frequency of the GWB. The bounds on the effective number of species from BBN/CMB are $\Delta N_{\rm{eff}} \lesssim 0.2$~\cite{Planck:2018jri, MANGANO2011296, RevModPhys.88.015004, peimbert2016primordialheliumabundancenumber}. We take $f_{\rm GW}^{\rm min} \sim 10^{-15} \ \rm{Hz}$\footnote{As is discussed in~\cite{Smith_2006, Cabass_2016}, the IR limit will be the horizon size at CMB decoupling ($\sim 10^{-17} \ \rm{Hz}$). However, we take the cut-off to be of order $10^{-15}$ Hz, since, realistically, the gravitational waves will need to oscillate for a while after entering the horizon before contributing to $\rho_{\rm rad}$.}, and $f_{\rm GW}^{\rm max}$ is given by Eq.~\ref{eq: k max}.


\section{Results}\label{sec: results}

Here, we present the results of our analysis. The results are shown considering various constraints and theoretical assumptions, and with progressively increasing dimensionality of the parameter space of primordial GWs with quantum decoherence effects. This allows us to study the specific effects on the parameter space, leading to a better understanding of the underlying physics. 
Our aim is to provide, for the first time, data-driven allowed regions for the decoherence parameter space. 
Given that all parameters are compatible with zero, we expect to find vast flat regions, where there is no preferred value.
Of course, it is challenging to obtain a perfectly flat distribution of posterior points in those regions with a finite number of samples. Indeed, there will always be under-sampled regions of parameters that could give the illusion of an actual disfavor for certain values of the parameters. For this reason, in the figures below, we apply some additional smoothing to the results for recovering flatness and avoiding over-interpreting the posterior plots.\footnote{In \texttt{getdist}, this can be controlled through \texttt{smooth\_scale\_1D} = 0.4 and \texttt{smooth\_scale\_2D} = 0.4.}

\subsection{Starobinsky}\label{sec: results Starobinsky}

\begin{figure}
    \centering
    \begin{subfigure}{0.48\textwidth}
        \centering
        \includegraphics[width=\linewidth]{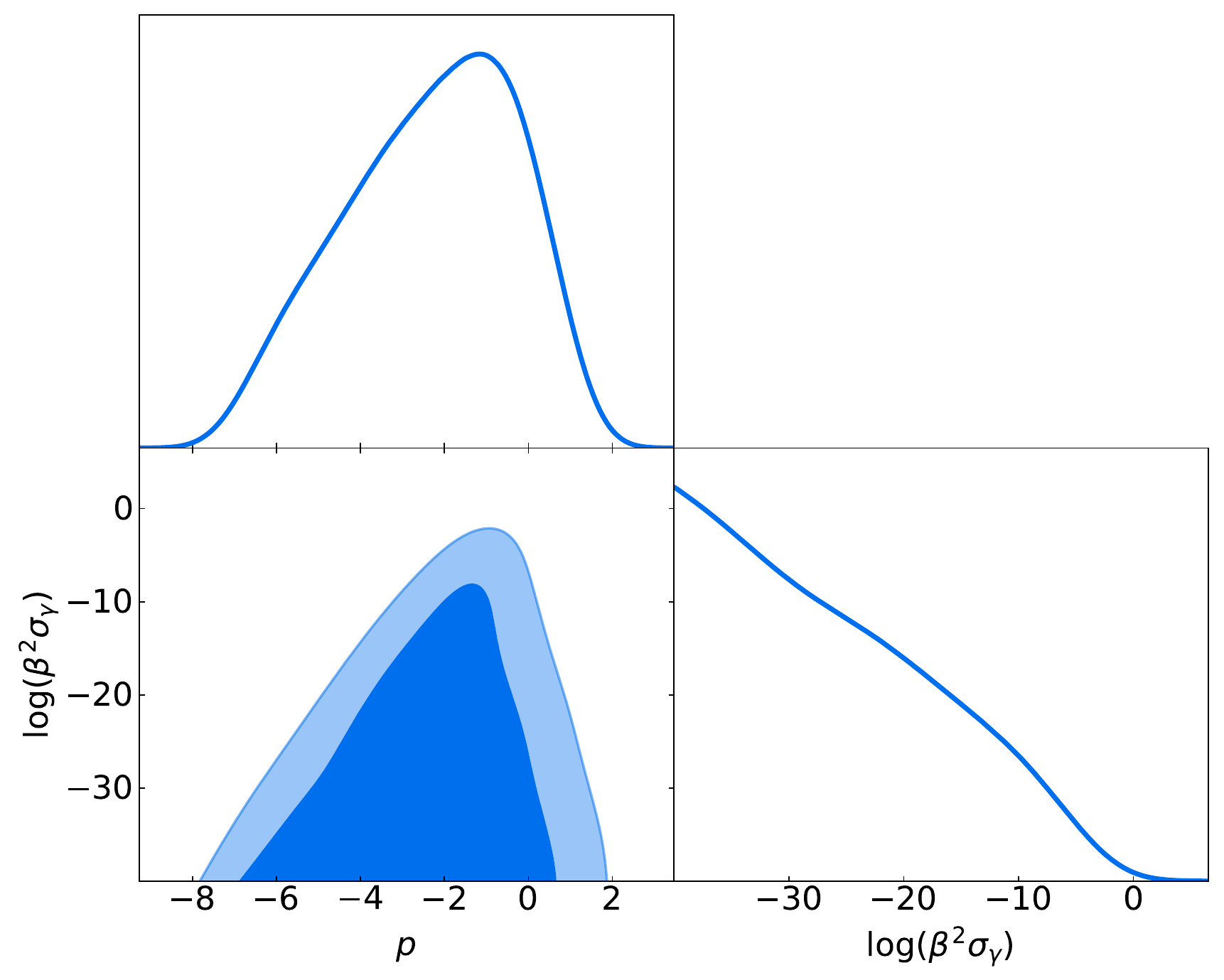}
        \label{fig: p sigma}
    \end{subfigure}
    \begin{subfigure}{0.48\textwidth}
        \centering
        \includegraphics[width=\linewidth]{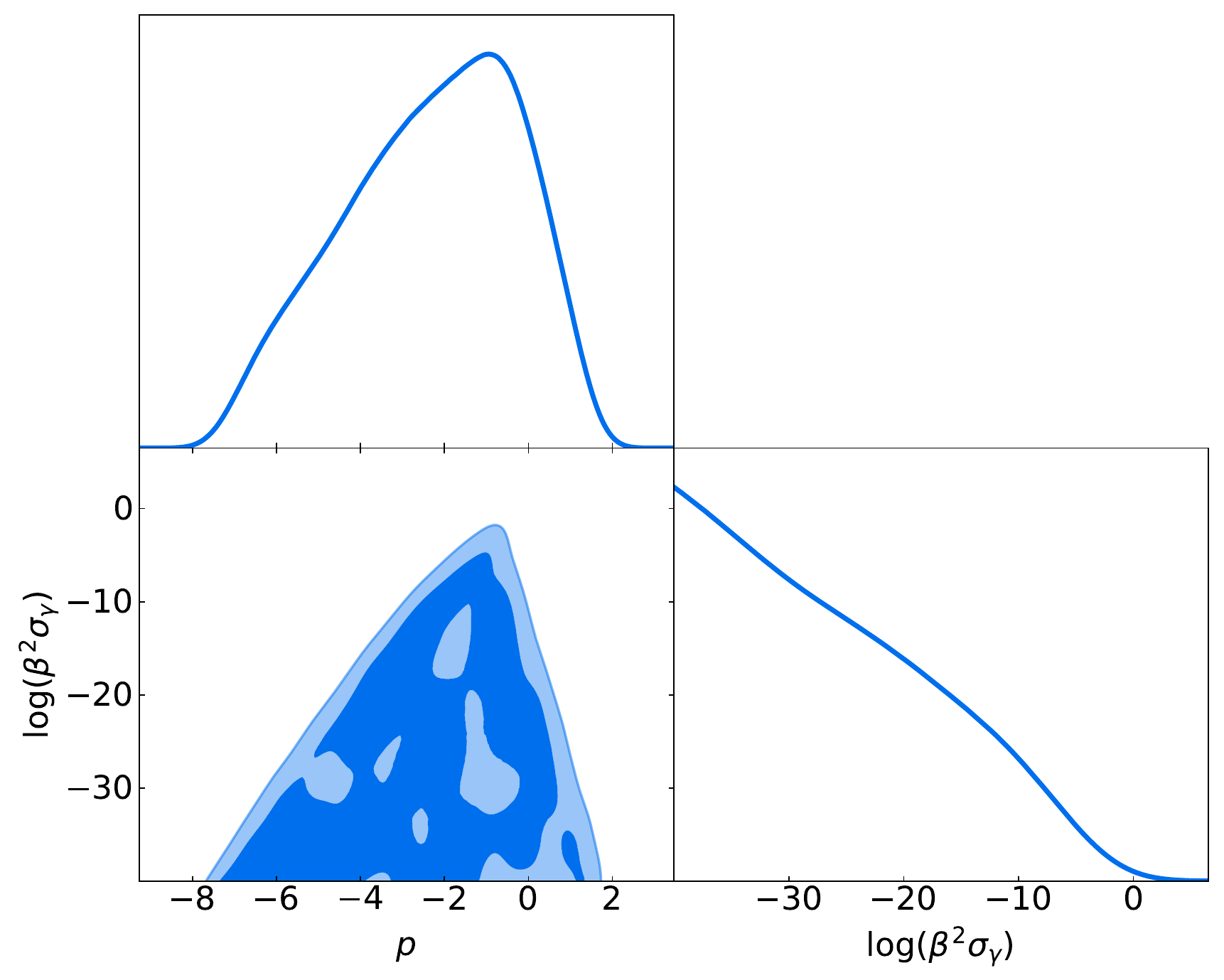}
        \label{fig: unsmooth}
    \end{subfigure}
    \caption{The 1D and 2D 68\% and 95\% CL interval constraints on $p, \beta^2\sigma_\gamma$, using PL+BK18+LVK. We set $r_{\ast}=0.00461$ to the value of Starobinsky model of inflation \cite{STAROBINSKY198099}. Additionally, we set $n_T=0, H_{\ast}l_E = 10^{-3}$, and $\Delta N_{\ast} = 50$. In the left panel, we apply a smoothing, namely \texttt{smooth\_scale\_1D = 0.4}, and \texttt{smooth\_scale\_2D = 0.4}. In the right panel, we apply no such smoothing to highlight that the smoothing does not affect the conclusions drawn in this work.}
    \label{fig: p sigma both}
\end{figure}

Here, we start by highlighting how $p$ and $\beta^2\sigma_\gamma$ are related to each other, while keeping the tensor-to-scalar ratio fixed to the Starobinsky model \cite{STAROBINSKY198099} prediction, namely the observational goal of future CMB surveys $r_{\ast}=0.00461$ \cite{LiteBIRD:2022cnt}. The results are presented in Fig.~\ref{fig: p sigma both}, where we also show the comparison between smoothed and unsmoothed posteriors, in the left and right panels, respectively. As mentioned above, we apply some additional smoothing to avoid over-interpreting undersampled regions. The figure demonstrates the effect described above and illustrates that the smoothing has not altered the results in any substantial way.

We find 
\begin{equation}
    \log(\beta^2\sigma_\gamma) < -7.9 \ \ (95\% \ \rm{CL}).
\end{equation}
From Fig.~\ref{fig: p sigma both} it is clear that for the highest values of $\beta^2\sigma_\gamma$, only a small region around $p=-1$ is allowed. These scenarios correspond to a (nearly) scale-invariant tensor power spectrum, where even a small quantum decoherence contribution already saturates the existing bound. The 1D distribution of $\beta^2\sigma_\gamma$ is prior-dominated for small values, due to the artificially imposed lower limit of $\beta^2\sigma_\gamma = 10^{-40}$. In principle, there is no lower limit for $\beta^2\sigma_\gamma$; thus, if the prior were to be extended to lower values, the posterior seen for the 2D distribution would continue downward indefinitely.

For $p$, we find 
\begin{equation}
    -6.4 < p < 1.2 \ \ (95\% \ \rm{CL}).
\end{equation}
However, for the reason explained above, this interval is strongly dependent on the chosen prior for $\beta^2\sigma_\gamma$. In other words, for $\beta^2\sigma_\gamma$ extremely small, $p$ is completely unconstrained by data; thus, this bound will only be used in the following to compare the constraining power on $p$ between different configurations.

Note that the allowed region for $p > -1$ is much smaller than the region for $p<-1$. This is because extremely blue-tilted scenarios are tightly constrained by LVK on very small scales. The scenarios with the highest tilt allowed correspond to a value for $\beta^2\sigma_\gamma$ such that the decoherence contribution to the power spectrum becomes dominant on scales outside of our observational window. 
These results are consistent with the estimate obtained in~\cite{de_Kruijf_2024}.

\subsection{Varying \texorpdfstring{$r, p, \beta^2\sigma_{\gamma}$}{TEXT}}\label{sec: results r, p, sigma}

\begin{figure}
    \centering
    \includegraphics[width=0.8\linewidth]{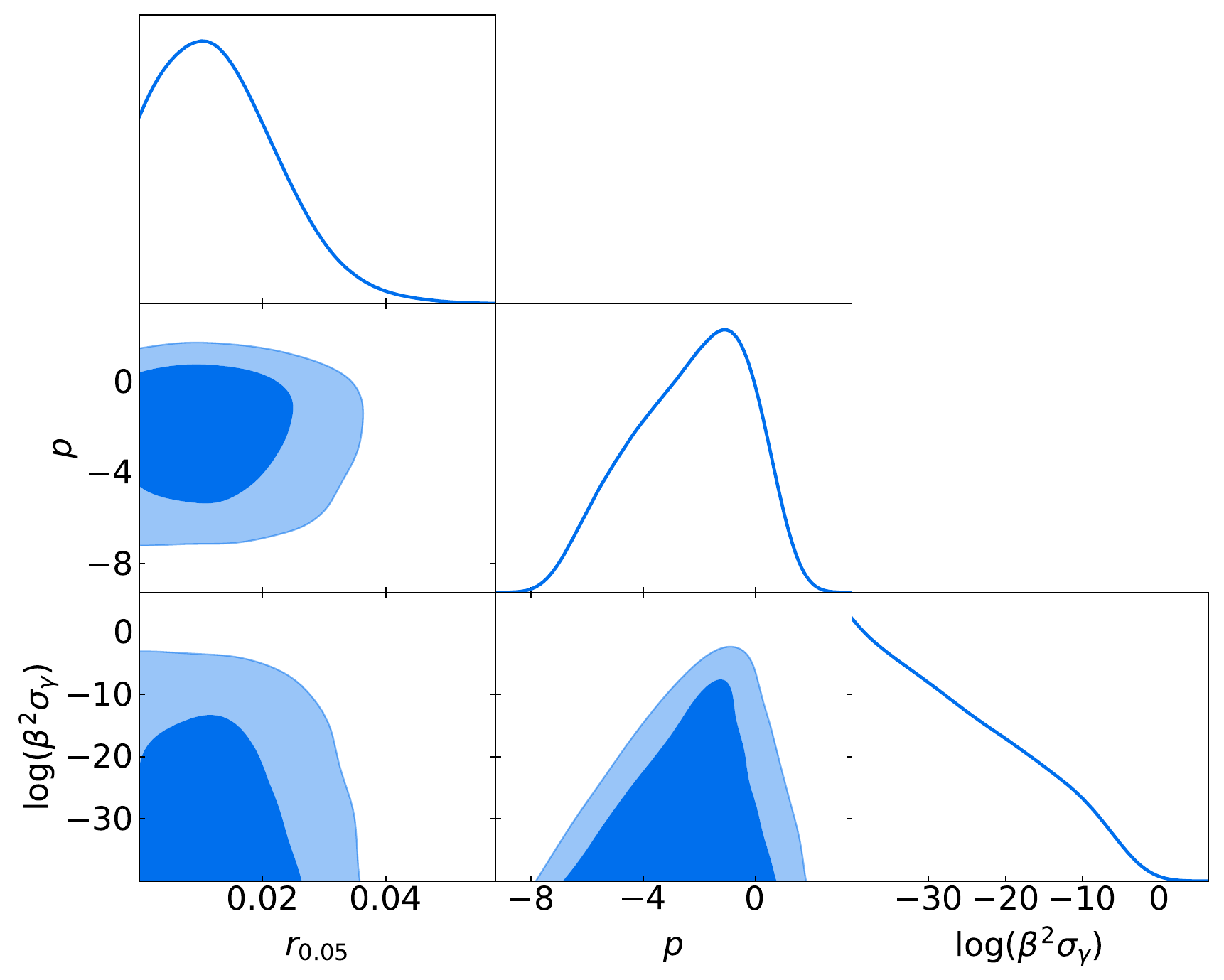}
    \caption{The 1D and 2D 68\% and 95\% CL interval constraints on $r_{0.05}, p, \beta^2\sigma_\gamma$, using PL+BK18+LVK. Additionally, we set $n_T=0, H_{\ast}l_E = 10^{-3}$, and $\Delta N_{\ast} = 50$. }
    \label{fig:r_p_sigma}
\end{figure}

In this section, we allow $r$ to vary, as shown in Fig.~\ref{fig:r_p_sigma}. The upper bound we find is 
\begin{equation}
    r_{0.05} < 0.031 \ \ (95\% \ \rm{CL}),
\end{equation} 
consistent with~\cite{Tristram_2022}, corresponding to the scenarios where the decoherence contribution is negligible on CMB scales.  
For high values of $r_{0.05}$, we see that only small values of $\beta^2\sigma_\gamma$ are allowed, since this means the decoherence contribution becomes dominant on scales much smaller (or much larger) than the pivot scale. Furthermore, the largest values of $r_{0.05}$ are allowed for $p\simeq-1$, corresponding to the scale invariant scenario. This is because, for this value of $p$ and small values of $\beta^2\sigma_\gamma$, the decoherence contribution is subdominant on all scales. At the same time, when $r_{0.05}$ is small, large values of $\beta^2\sigma_\gamma$ can be accommodated, just like scenarios where $p$ varies from $-1$ significantly.

For low values of $r$, a broader range of the other parameters is allowed. However, there is still a large preference for small values of $p$, due to the LVK constraints on very small scales, allowing only for a small effective spectral tilt. As $r_{0.05}$ becomes small, the allowed region for $\beta^2\sigma_\gamma$ quickly converges to its maximum allowed value, corresponding to the decoherence contribution becoming dominant on scales outside of our observational window. The upper bound we find is 
\begin{equation}
    \log(\beta^2\sigma_\gamma) < -7.8 \ \ (95\% \ \rm{CL}),
\end{equation}
which is compatible with the previously obtained results.

Finally, in this scenario, the $p$ interval is given by 
\begin{equation}
    -6.4 < p < 1.2 \ \ (95\% \ \rm{CL}).
\end{equation} 
This is the same range as for the Starobinsky case considered above, showing that allowing $r_{0.05}$ to vary has no significant impact on our results. The behavior seen here is the same as before, with a large freedom for red-tilted power spectra, but tighter constraints for strongly blue-tilted spectra due to the small-scale upper limit. Of course, this is for the $\beta^2\sigma_\gamma$ range considered in this work. For smaller values of $\beta^2\sigma_\gamma$, a broader range of $p$ will be allowed.

\subsection{Decoherence criterion}\label{sec: results decoherence criterion}

An interesting scenario to explore is the one where we assume the tensor modes on CMB scales have fully decohered. It is shown in~\cite{de_Kruijf_2024} that using the decoherence criterion on these scales provides a lower bound on the combination of $p$ and $\beta^2\sigma_\gamma$.
To do so, we define a simple step-likelihood on $\delta_{\boldsymbol{k}_\ast}$, as a function of $p$ and $\beta^2\sigma_\gamma$, which rejects sampled points with $\delta_{\boldsymbol{k}_\ast} < 1$ (see Eq.~\eqref{eq: decoherence criterion}). 
This leads to the upper bound 
\begin{equation}
    \log(\beta^2\sigma_\gamma) < -4.8 \ \ (95\% \ \rm{CL}),
\end{equation}
being higher than for the scenarios considered previously, due to low values of $\beta^2\sigma_\gamma$ now being excluded. 
However, the upper bound 
\begin{equation}
    r< 0.031 \ \ (95\% \ \rm{CL}).
\end{equation} 
remains the same. 

In Fig.~\ref{fig: r_p,_sigma_dec_crit}, the 2D posterior on the $\log(\beta^2\sigma_\gamma)-p$ plane shows a clear anti-correlation, where an increase in $\beta^2\sigma_\gamma$ corresponds to a decrease in $p$.
This correlation is obvious from Eq.~\eqref{eq: decoherence criterion}, when enforcing $\delta_{\boldsymbol{k}_\ast} > 1$. It is clear that imposing this lower limit severely restricts the amount of $p$ values allowed, with respect to Fig.~\ref{fig:r_p_sigma}. 
As is further evidenced by the preferred range, namely,
\begin{equation}
    -1.4 < p < 1.5 \ \ (95\% \ \rm{CL}).
\end{equation}
This shows significant changes with respect to the previous cases. Nearly all of the red-tilted scenarios ($p<-1$) are disallowed, showing that these correspond to low values of $\beta^2\sigma_\gamma$. Due to the decoherence criterion, we see that the lower bound on $p$ is no longer dependent on the $\beta^2\sigma_\gamma$ prior.
Furthermore, slightly more blue-tilted scenarios are allowed now, where Fig.~\ref{fig: r_p,_sigma_dec_crit} shows a preference for lower values of $\beta^2\sigma_\gamma$.

\begin{figure}
    \centering
    \includegraphics[width=0.8\linewidth]{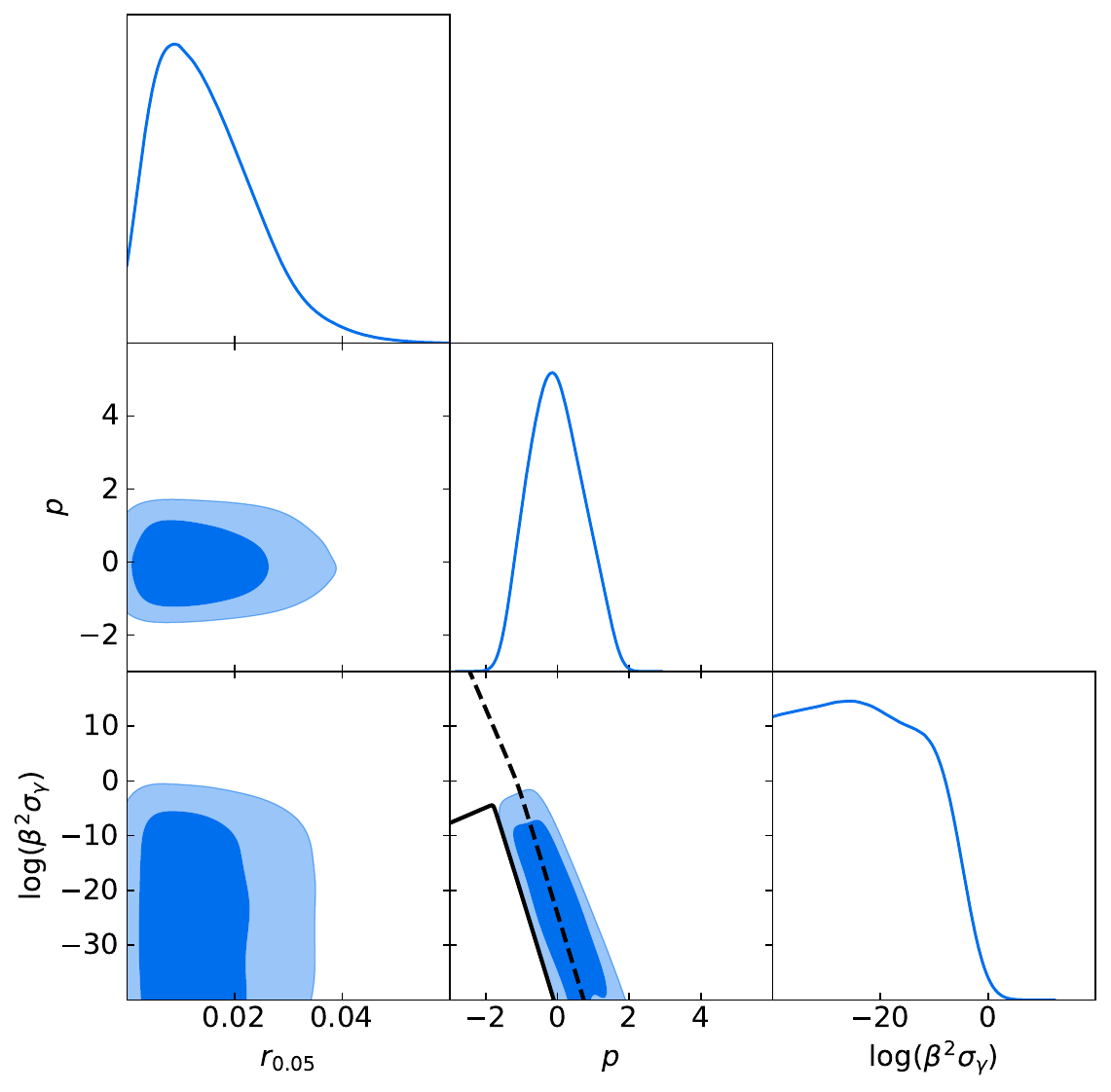}
    \caption{The 1D and 2D 68\% and 95\% CL interval constraints on $r_{0.05}, p, \beta^2\sigma_\gamma$, using PL+BK18+LVK. We also enforce the decoherence criterion on CMB scales, shown as the black solid line. Additionally, the dashed black line indicates the values of $p$ and $\log(\beta^2\sigma_\gamma)$ where $\delta_{\boldsymbol{k}_{\rm LVK}} =1$, meaning that above this line the scales LVK probes have fully decohered (this is not implemented in this work). Furthermore, we set $n_T=0, H_{\ast}l_E = 10^{-3}$, and $\Delta N_{\ast} = 50$.}
    \label{fig: r_p,_sigma_dec_crit}
\end{figure}

To highlight the significance of the decoherence criterion, in Fig.~\ref{fig: r_p,_sigma_dec_crit}, we have also added the line $\delta_{\boldsymbol{k}_{\rm LVK}}=1$. This indicates the completion of the quantum decoherence on the scales probed by LVK. It provides us with a new potential lower bound on the values of $p$ and $\log(\beta^2\sigma_\gamma)$. For similar lines corresponding to other scales, please see Fig. 5 of~\cite{de_Kruijf_2024}. If we impose that even these small LVK scales have fully decohered, only the region above the black line is observationally allowed. This significantly decreases the allowed parameter space, meaning that to achieve this (for a given interaction strength), we need an even steeper time dependence of the interaction. Clearly, in order for all scales to have decohered completely, we either need a very strong interaction or the interaction needs to sharply increase over time. On the other hand, let us stress that these are not necessarily the most likely scenarios, meaning it is possible that on small scales the system has not fully decohered, and on those scales quantum signatures survive. 

Nevertheless, let us reiterate that the decoherence criterion as a theoretical assumption has some caveats. Namely, it has not been proven that the inflationary GWB is classical on CMB scales, and we assume the CMB modes are only decohered through the interaction focused on in this work.

\subsection{Tilted power spectrum}

In the previously considered scenarios, we assume the ``standard'' GW power spectrum is scale-invariant ($n_T=0$). 
However, in this section, we consider a primordial power spectrum with an inherent spectral tilt ($n_T \neq 0$), below referred to using the word standard. 
Since quantum decoherence changes the power spectrum into a broken power-law, it is clear that introducing a standard tilt has a similar effect on the total primordial power spectrum as some of the quantum decoherence scenarios considered. Especially since we only have observational upper bounds on the GW power spectrum, we do not have access to its full shape.

\begin{figure}
    \centering
    \includegraphics[width=0.9\linewidth]{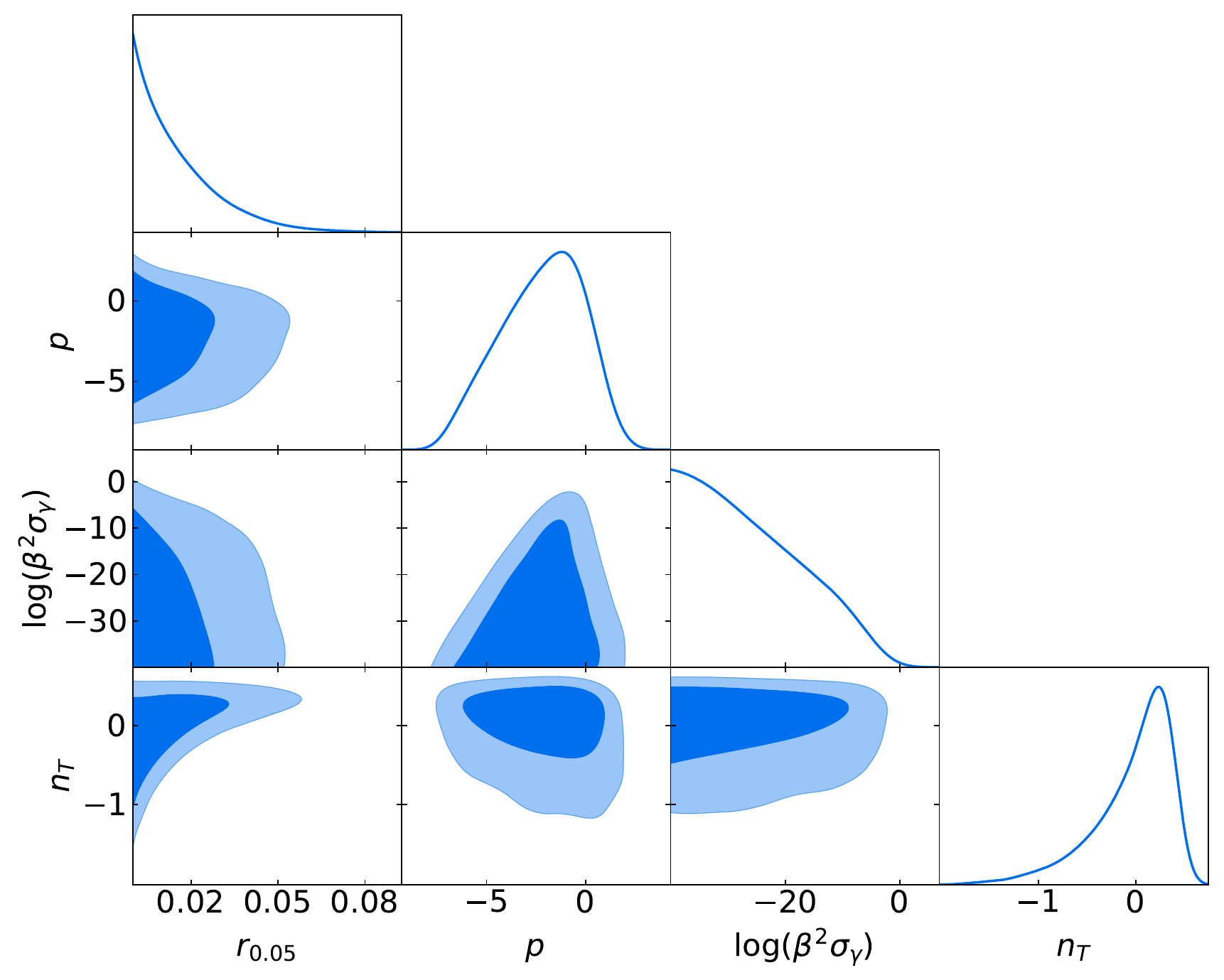}
    \caption{The 1D and 2D 68\% and 95\% CL interval constraints on $r_{0.05}, p, \beta^2\sigma_\gamma, n_T$, using PL+BK18+LVK. Additionally, we set $H_{\ast}l_E = 10^{-3}$, and $\Delta N_{\ast} = 50$.}
    \label{fig:tilted PS}
\end{figure}

As can be seen from Fig.~\ref{fig:tilted PS}, only a small range of tilts is allowed by observations, with a clear preference for a tilt around $n_T = 0$. The bounds we find on the standard power-law parameters correspond to 
\begin{equation}
    r_{0.05}< 0.045, \ -0.9 < n_T < 0.6 \ \ (95\% \ \rm{CL}).
\end{equation}
The upper bound on $r_{0.05}$ is significantly higher than found in previous results. This, however, is due to the additional freedom on $n_T$, rather than on the decoherence parameters.
To understand this, we can compare this with the literature \cite{Galloni_2023}; however, this is not trivial, as the tensor-to-scalar ratio is evaluated at $0.05$ Mpc$^{-1}$ in this work. Indeed, $r$ is typically evaluated at $0.01$ Mpc$^{-1}$ when sampling $n_T$, since that is the de-correlation scale of those two parameters \cite{Planck:2018jri}. Thus, in Appendix~\ref{sec: r0.01} we repeat this analysis on the $r_{0.01}-n_T$ plane. Similarly to Section~\ref{sec: results r, p, sigma}, we show that sampling the decoherence parameters has no significant impact on the tensor parameter. In other words, this demonstrates that the increase in the upper bound of $r_{0.05}$ is only due to the extra freedom in the value of $n_T$ and not of the decoherence parameters.

The bounds on the decoherence parameters are given by 
\begin{equation}
    \log(\beta^2\sigma_\gamma) < -8.0, \ -6.4 < p < 1.4 \ \ (95\% \ \rm{CL}).
\end{equation} 
These results are very similar to those with $n_T=0$. This is not surprising, since Fig.~\ref{fig:tilted PS} shows that a large part of the parameter space corresponds to low values of $r_{0.05}$ and a weakly tilted standard power spectrum ($n_T$ small).

Furthermore, the 2D contours on $n_T$ and $p$ behave as expected, since for red-tilted standard power spectra ($n_T<0$), we cannot have a strong red-tilted decoherence contribution, as to avoid the upper bound set by observations on CMB scales. The same logic applies to blue-tilted scenarios, constrained on LVK scales. Finally, it is clear from the 2D constraints shown in Fig.~\ref{fig:tilted PS}, that any strong standard tilt corresponds to low values in $\beta^2\sigma_\gamma$. This means that for very tilted standard scenarios, the decoherence contribution becomes dominant on scales further away from the pivot scale. This leads to a smaller increase in the total power spectrum and therefore does not conflict with the observational upper bounds.

The observed relations between the standard and decoherence parameters highlight the importance of considering quantum decoherence when studying the GW power spectrum. As of now, it is not possible to distinguish whether a potentially observed tilt of the total GW power spectrum is due to a standard spectral tilt or due to the contribution from quantum decoherence. However, the presence of a broken power-law GW power spectrum could point towards the quantum decoherence scenarios considered in this paper.
To confirm this, further research into quantum decoherence during inflation and its effects on, e.g. higher order correlation functions, needs to be done. We leave this for future work.

\subsection{Full parameter space}\label{sec: full par space}

In this section, we open up the full parameter space, taking into account more general models of inflation (varying $\Delta N_\ast$) and the correlation length of the environment (varying $H_\ast l_E$). In opening up the parameter space, we allow for a cut-off due to $k_{\rm max}$ (see Section~\ref{sec: tensor power spectrum} for more details),  to be on scales larger than those probed by LVK.
This means we could have an increase due to quantum decoherence on scales smaller than the CMB, which then stops on some intermediate scale, larger than those probed by LVK.
We study these scenarios both with and without taking the recent NANOGrav detection into account, allowing us to see exactly how NANOGrav helps to constrain these scenarios. Furthermore, we add the upper bound provided by BBN constraints as well, giving constraints on very high frequencies.

\begin{figure}
    \centering
    \includegraphics[width=1.0\linewidth]{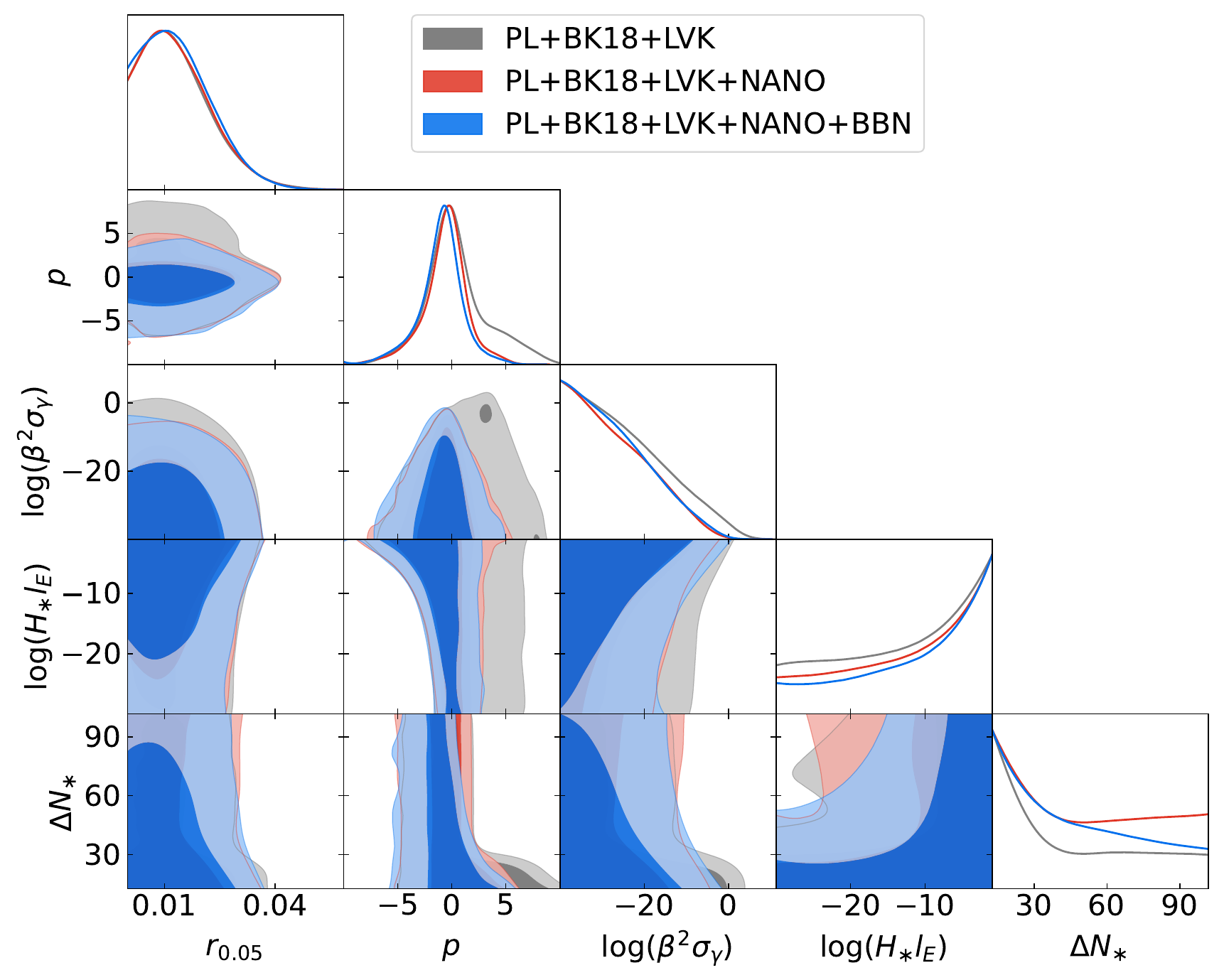}
    \caption{The 1D and 2D 68\% and 95\% CL interval constraints on $r_{0.05}, p, \beta^2\sigma_\gamma, H_\ast l_E, \Delta N_\ast$. The results are shown using PL+BK18+LVK (grey), PL+BK18+LVK+NANO (red), and PL+BK18+LVK+NANO+BBN (blue). Additionally, we set $n_T=0$. We have negated any smoothing to highlight the peaked scenarios on intermediate scales, allowed by PL+BK18+LVK.}
    \label{fig: full par}
\end{figure}

Fig.~\ref{fig: full par} shows the constraints on the full parameter space, for three combinations of constraints. Let us start by discussing the results obtained using only the CMB and LVK. First, we find that the upper bound on $r_{0.05}$ remains the same when opening up the full parameter space, namely,
\begin{equation}
    r_{0.05} < 0.031 \ \ (95\% \ \rm{CL}),
\end{equation} 
consistent with the literature~\cite{Tristram_2022}. However, for any decoherence scenario corresponding to a strong increase in the power spectrum, namely $p$ deviating strongly from $-1$, large values of $\beta^2\sigma_\gamma$ or small values of $Hl_E$, a lower value of $r_{0.05}$ is preferred, as is clear from their respective 2D contours.

Secondly, the 1D posterior of $ \log(Hl_E)$ in Fig.~\ref{fig: full par} shows a preference for high values, although the posterior plateaus away from the upper bound of the prior.
It is clear from the 2D distribution in  Fig.~\ref{fig: full par} that for almost all values of $\Delta N_\ast$, the observations prefer a high value of $H_\ast l_E$. 
This observational preference seems in contrast with the assumption $Hl_E \ll 1$ imposed in the derivation. 
The integrals in the power spectrum definition (Eq.~\eqref{eq:I1,I2 definition}) provide a potential explanation.
For red-tilted power spectra ($p<-1$), the interaction strength decreases with time, meaning it is strongest at early times. Now, if $H_\ast l_E$ is very small, the modes cross out of the environmental correlation length at an earlier time, when the interaction is stronger, leading to a larger increase in the power spectrum than for higher values of $H_\ast l_E$. Meaning that most red-tilted scenarios with very small values of $Hl_E$ are disallowed. 
However, for the blue-tilted scenarios, this is not the case. The interaction is strongest at later times, meaning that the modes are mainly affected by the environment at later times as well, making this effect less noticeable. On the other hand, the upper integration limits of the integrals of Eq.~\eqref{eq:I1,I2 definition} are given by the inverse of $(H_\ast l_E)$, meaning that a smaller environmental correlation length leads to a larger integration interval and therefore a stronger increase in the power spectrum. This, in turn, is observationally disallowed by LVK. If possible UV contributions to the power spectrum were to be considered, the cut-off would be dependent on $H_\ast l_E$, and therefore we could see a stronger imprint. However, in this work we enforce the cut-off due to $k_{\rm max}$, leaving $H_\ast l_E$ only constrained through it's affect on red-tilted power spectra.

Furthermore, in the 1D posterior, there seems to be a slight preference for shorter periods of inflation. Although the constraints are again prior-dependent, imposed based on the theoretical bounds found in~\cite{Liddle_2003}.
For a very short period of inflation, $\Delta N_{\ast} \sim 20$, we see that the rest of the parameter space opens up more. This is reflected in the upper bound 
\begin{equation}
    \log(\beta^2\sigma_\gamma) < - 4.9  \ \ (95\% \ \rm{CL}),
\end{equation}
which is higher than the bounds found before, when considering a smaller parameter space. 
This is because for those values, the cut-off scale of the power spectrum is on scales smaller than those probed by LVK.
This means it is allowed to have a peak between CMB and LVK scales, freeing up the parameter space for strong tilted power spectra (large values of $p$) and those with an increase starting on large scales (large values of $\beta^2 \sigma_\gamma$), as is clearly visible from the grey bubble seen in their 2D interval region.
The larger freedom is also evident when considering the bounds on the decoherence-induced spectral tilt, namely,
\begin{equation}
    -5.3 < p < 7.3 \ \ (95\% \ \rm{CL}).
\end{equation} 
The increase in large values of $p$ allowed, with respect to previous results, is considerable. This reflects again the increased freedom when considering a bump on intermediate scales. In conclusion, the perceived preference for $\Delta N_\ast$ is driven by the projection effects caused by the possible bump in the power spectrum on intermediate scales. This parameter should be considered completely unconstrained.

Then, we include the NANOGrav detection as an upper bound (using $\mathcal{L}_{\rm NANO}^{\rm bound}$), and see that some of the most extreme scenarios are no longer allowed. The general constraints on $r_{0.05}, H_\ast l_E, \Delta N_\ast$, as shown in Fig.~\ref{fig: full par}, remain quite the same. Although the scenarios corresponding to an extreme bump on intermediate scales are disallowed, since the bump cannot exceed the NANOGrav upper bound. This is mainly the parameter space with the highest spectral tilt ($p$), and the highest values of $\beta^2\sigma_\gamma$. As is reflected in the bounds
\begin{equation}
    \log(\beta^2\sigma_\gamma) < - 10.3, \ -5.7 < p < 3.8 \ \ (95\% \ \rm{CL}).
\end{equation} 
These are quite drastic changes with respect to the previous results, showing exactly the type of scenarios disallowed by the NANOGrav results. Furthermore, we see that imposing this upper bound significantly changes the 1D distribution of $\Delta N_\ast$. There are no clear bounds to be found on this parameter. Higher values ($\Delta N_\ast > 45$), corresponding to more standard models of inflation, are more preferred than before, although the distribution is still skewed to it's lowest values. The bounds on $\log(Hl_E)$ remain practically unchanged, with a preference for higher values.

Finally, we add the upper bound on $\Omega_{\rm GW}$ from BBN. 
At high frequencies ($> 10$ kHz)~\cite{Servant_2024, 2021LRR....24....4A, 2018CQGra..35p3001C}, this provides constraints competitive with the other observables considered in this work, meaning it helps constrain the scenarios with the strongest blue-tilt. As can be seen from Fig.~\ref{fig: full par}, adding the BBN constraints further disallows the regions with the highest blue-tilt, giving 
\begin{equation}
    -6.0 < p < 3.2 \ \ (95\% \ \rm{CL}).
\end{equation}
It also shows a preference for a shorter period of inflation, since this lowers $f_{\rm GW}^{\rm max}$ and therefore the integration interval of Eq.~\eqref{eq: BBN bound}. However, it slightly affects the upper bound 
\begin{equation}
    \log(\beta^2\sigma_\gamma) < - 9.8  \ \ (95\% \ \rm{CL}),
\end{equation} although the difference with the previous results (PL+BK18+LVK+NANO) is too small to be significant for the parameter range considered in this work. 
Adding the BBN bound, in combination with the CMB, LVK, and NANOGrav, we do recover similar behavior for both $\Delta N_\ast$ and $\log(Hl_E)$ as before, both tending to smaller and higher values, respectively. This is also reflected in the change of the 1D distribution function of $\Delta N_\ast$, showing a decreased preference for high values, tending more towards the CMB+LVK results.

\begin{figure}
    \centering
    \includegraphics[width=\linewidth]{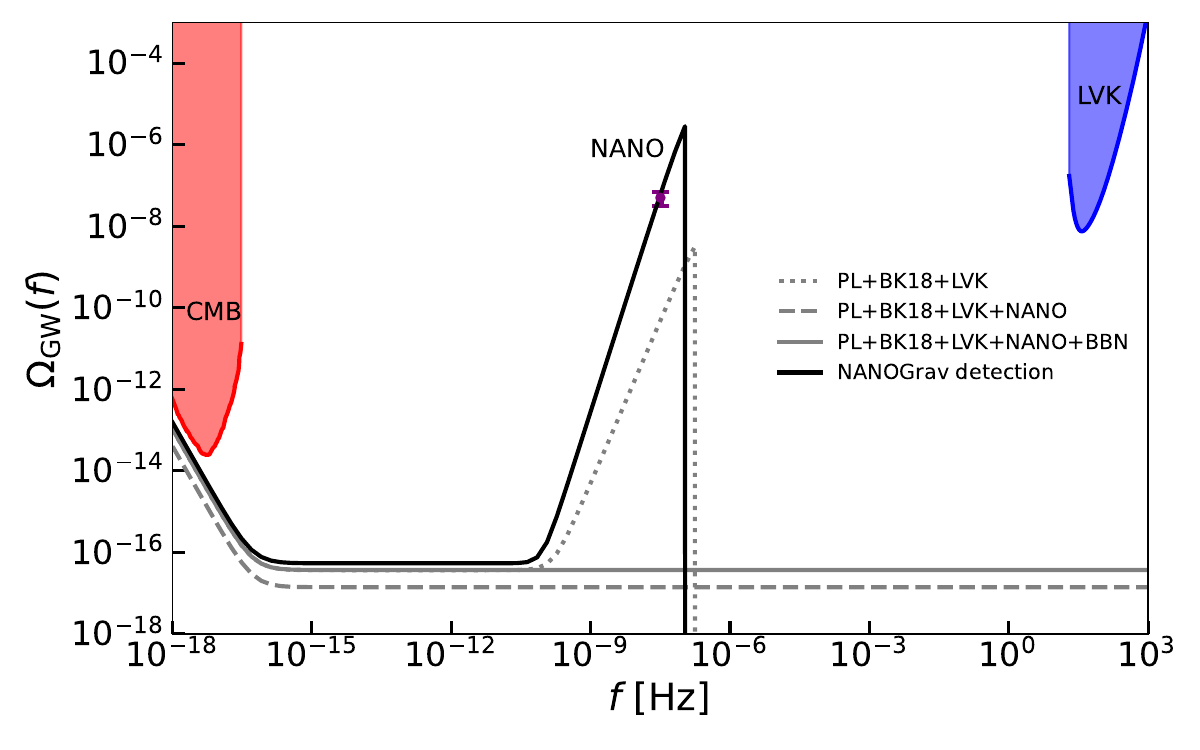}
    \caption{The best fit for the GW energy density $\Omega_{\rm GW}$ as a function of frequency, shown in dotted, dashed, and solid grey for the  full parameter space analysis using the PL+BK18+LVK, PL+BK18+LVK+NANO, PL+BK18+LVK+NANO+BBN constraints, respectively. The solid black line shows the best fit values found corresponding to the NANOGrav detection. 
     for a variety of quantum decoherence scenarios corresponding to the NANOGrav detection (shown in purple).
    Furthermore, we show the constraints from CMB measurements at low multipole moments in red, the NANOGrav detection in purple, and the upper bound from LVK in blue~\cite{Abbott_2021, Thrane_2013, 2023ApJ...951L...8A}. }
    \label{fig: Omega all scenarios}
\end{figure}

To illustrate the above-mentioned results and the type of power spectra quantum decoherence induces, we show the best fit for several scenarios considered above in Fig.~\ref{fig: Omega all scenarios}. The dotted, dashed, and solid grey line correspond to the best fit obtained for the full parameter space analysis using the PL+BK18+LVK, PL+BK18+LVK+NANO, and PL+BK18+LVK+NANO+BBN constraints, respectively. This clearly shows the cut-off of the inflationary GW power spectrum.
These best fits reflect the preference for a power spectrum without a bump on intermediate scales, when adding the NANOGrav detection as an upper bound, even though the best fit scenario found without it would still be allowed.

\subsection{NANOGrav detection}

Finally, the origin of the NANOGrav detection is currently debated. Therefore, in addition to using the NANOGrav detection as an upper bound on the GWB, it is interesting to ask ourselves which quantum decoherence scenarios would correspond to the signal observed with NANOGrav (using $\mathcal{L}_{\rm NANO}^{\rm det.}$). Indeed, for the GW power spectrum to correspond to the NANOGrav signal, while staying within the upper bounds set by CMB and LVK, we need the peak behavior mentioned above on intermediate scales. This is shown in Fig.~\ref{fig: Omega all scenarios}, where the black line shows the best fit for the NANOGrav detection scenario. In this case, the LVK upper bound is also imposed, forcing the cut-off of the power spectrum on intermediate scales (for more details on this see Section~\ref{sec: tensor power spectrum}).

\begin{figure}
    \centering
    \includegraphics[width=1.0\linewidth]{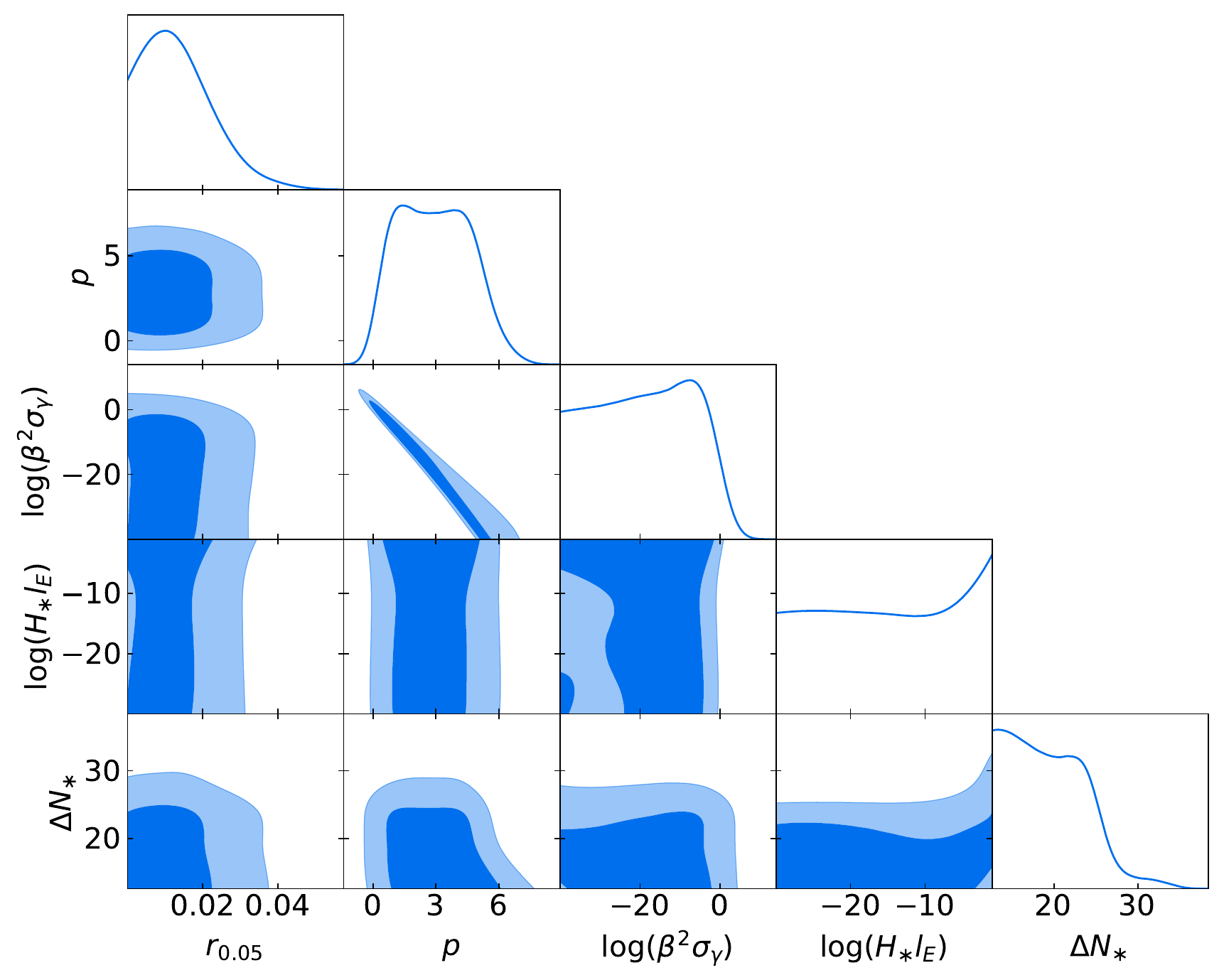}
    \caption{The 1D and 2D 68\% and 95\% CL interval constraints on $r_{0.05}, p, \beta^2\sigma_\gamma, H_\ast l_E, \Delta N_\ast$, using PL+BK18+LVK and imposing the NANOGrav detection. Additionally, we set $n_T =0$.  }
    \label{fig:nano detection}
\end{figure}

The region of the parameter space able to explain the NANOGrav detection is shown in Fig.~\ref{fig:nano detection}, corresponding to a very specific range of the quantum decoherence parameter space. Most importantly, we require very short inflation, namely 
\begin{equation}
    \Delta N_\ast < 26.4 \ \ (95\% \ \rm{CL}),
\end{equation}
which can give a peak on intermediate scales.
Moreover, we need a very specific combination of the tilt $p$ and interaction strength $\beta^2 \sigma_\gamma$, to ensure the correct increase at the NANOGrav frequency range. This correlation can be approximated by the linear fit $\log(\beta^2\sigma_\gamma) = -7.0 p + 1.4$, for the bounds 
\begin{equation}
    \log(\beta^2\sigma_\gamma) < -0.6, \ -0.1 < p < 6.0\ \ (95\% \ \rm{CL}).
\end{equation} 
However, let us note that in this work we consider the NANOGrav results modeled with a variable power-law exponent, at a single frequency ($f_{\rm yr} = 1 \mathrm{yr}^{-1}$). We do not fit the full shape of our power spectrum to the NANOGrav data, which could limit the decoherence spectral tilt allowed ($p$), this is left for future work. 
Finally, the NANOGrav detection does not provide any clear preference for a value of $Hl_E$, which is due the fact that in this case the cut-off of the power spectrum is due to $k_{\rm max}$. However, as is described in Section~\ref{sec: tensor power spectrum}, even without explicitly enforcing $k_{\rm max}$, the environmental correlation length will provide a cut-off, and for high values of $H_\ast l_E$ this cut-off approaches $k_{\rm max}$.
Moreover, the upper limit on $r_{0.05}$ remains unchanged with respect to previous results, namely 
\begin{equation}
    r_{0.05} < 0.031\ \ (95\% \ \rm{CL}).
\end{equation}

To summarize, the bounds obtained in this work are shown in Table~\ref{tab: bounds results}. Let us reiterate that the bounds obtained on $p$ are dependent on the prior imposed on $\log(\beta^2\sigma_\gamma)$, meaning that for a much weaker interaction between the tensor modes and the environment scalar sector, a stronger time dependence of the interaction is observationally allowed.

\begin{table}
    \centering
    \begin{tabular}{|c|c|c|c|c|c|}
        \hline
        Constraints & $r_{0.05}$ & $\log(\beta^2\sigma_\gamma)$ & $p$ & $n_T $& $\Delta N_\ast$\\
        \hline
        Starobinsky & - & $<-7.9$ & $[ -6.4, 1.2]$ & - & - \\
        Varying $r, p , \beta^2\sigma_\gamma$ & $<0.031$ & $<-7.8$ & $[-6.4, 1.2]$ & - & - \\
        Decoherence criterion & $<0.031$ & $<-4.8$ & $[-1.1, 4.5]$ & - & - \\
        Tilted power spectrum & $<0.045$ & $<-8.0$ & $[-6.4, 1.4]$ & $[-0.9, 0.6]$ & - \\
        PL+BK18+LVK & $<0.031$ & $<-4.9$ & $[-5.3, 7.3]$ & - & - \\
        PL+BK18+LVK+NANO & $<0.031$ & $<-10.3$ & $[-5.7, 3.8]$ & - & - \\
        PL+BK18+LVK+NANO+BBN & $<0.031$ & $<-9.8$ & $[-6.0, 3.2]$ & - & - \\
        NANOGrav detection & $<0.031$ & $<-0.6$ & $[-0.1, 6.0]$ & - & $<26.4$ \\
        \hline
    \end{tabular}
    \caption{An overview of the observational bounds found in this work, for various observational and theoretical constraints, all obtained at 95\% CL. }
    \label{tab: bounds results}
\end{table}


\section{Conclusion}\label{sec: Conclusion}

During inflation, the interaction between the tensor modes and a (scalar) environment can lead to the quantum decoherence of the tensors. As studied in~\cite{de_Kruijf_2024}, this changes the GW power spectrum. In this work, this effect on the GW background is constrained using data from \textit{Planck}, BICEP/Keck array, LIGO-Virgo-KAGRA, NANOGrav, and BBN. In particular, we study the tensor-to-scalar ratio $r$, the strength of the interaction between the tensor modes and the scalar environment $\beta^2\sigma_\gamma$, the time dependence of that interaction $p$, the environmental correlation length $H_\ast l_E$, and the amount of $e$-folds between the pivot scale exiting the horizon and the end of inflation $\Delta N_\ast$.

In this work, we study various constraints and theoretical priors, with progressively increasing dimensionality of the parameter space of primordial GWs with quantum decoherence effects. An overview of the obtained bounds is shown in Table~\ref{tab: bounds results}. Here, we only summarize the results found when studying the full parameter space, while considering all observational constraints mentioned in this work.
First, the upper bound on the tensor-to-scalar ratio $r_{0.05}<0.031$ (95\% CL) is consistent with existing literature~\cite{Tristram_2022}, highlighting that when the decoherence contribution is subdominant, the parameter bounds remain unchanged. Second, the upper bound on the interaction strength is $\log(\beta^2\sigma_\gamma) < -9.8$ (95\% CL), showing that very strong interactions are observationally disallowed. 
Additionally, the allowed range for the time dependence of the interaction is $-6.0 < p < 3.2$ (95\% CL.), meaning that strong interactions that depend very strongly on time are disallowed. Note, however, that both lower and upper bounds on $p$ are strongly dependent on the prior range of $\log(\beta^2\sigma_\gamma)$ chosen in this work, as can be seen from their 2D contours. Thus, they are useful to compare this analysis with future works, but they should not be treated as strong observational bounds.

Moreover, we do not find a concrete upper or lower bound on the environmental correlation length $H_\ast l_E$ and the amount of $e$-folds between the pivot scale exiting the horizon and the end of inflation $\Delta N_\ast$. Although their posteriors, shown in Fig.~\ref{fig: full par}, indicate a preference for higher and lower values, respectively. Interestingly enough, these small $\Delta N_\ast$ and high $H_\ast l_E$ correspond to a cut-off of the GW power spectrum on intermediate scales (between NANOGrav and LVK scales), as can be seen in Section~\ref{sec: tensor power spectrum}.
However, in this work, we neglect any potential UV contribution due to quantum decoherence, meaning that the cut-off is only due to the small value of $\Delta N_\ast$. A more in depth study into the cut-off could lead to improved bounds on $H_\ast l_E$, since possibly a more realistic and detailed description of the cut-off would leave a stronger imprint. For now, the only constraints on $H_\ast l_E$ come from its effect on red-tilted power spectra (see the discussion in Section~\ref{sec: full par space}). However, for $\Delta N_\ast$ the preference for smaller values seems to be due to volume effects when allowing for scenarios with a bump on intermediate scales. This behavior is highlighted by the separate small grey region seen in the 2D contours between $\log(\beta^2\sigma_\gamma)$ and $p$, corresponding to extreme examples of such a bump. A considerable portion of these scenarios is disallowed when adding on the NANOGrav detection as an upper bound to the inflationary GW power spectrum, as is shown in red in Fig~\ref{fig: full par}.

However, these types of scenarios could also explain the NANOGrav detection, as is shown in Fig.~\ref{fig: Omega all scenarios}. The quantum decoherence scenarios that correspond to the signal found by NANOGrav are a small part of the parameter space. Specifically, with a linear correlation between $\log(\beta^2\sigma_\gamma)$ and $p$, a blue tilt $-0.1<p<6.0$, and a short period of inflation $\Delta N_\ast < 26.4$ (95\% CL.). This points to quantum decoherence as a plausible scenario that could explain the recent detection by PTA surveys. However, a goodness-of-fit analysis would be required to determine whether this model is actually favored over other possible sources. Such an analysis is beyond the scope of this work.

Finally, we study the effect of the duration of quantum decoherence on the observationally allowed scenarios. 
There is a clear correlation between the values of $p$ and $\beta^2\sigma_\gamma$ when enforcing that the largest scales (CMB) have fully decohered by the end of inflation, disallowing a large part of the parameter space (as seen in Fig.~\ref{fig: r_p,_sigma_dec_crit}). Namely, most of the red-tilted power spectra are disallowed ($-1.4 < p < 1.5$ at 95\% CL.), and the upper bound $\log(\beta^2\sigma_\gamma) < -4.8$ (95\% CL.) has shifted upwards with respect to other results found. 
However, interestingly enough, the fact that still a large part of the parameter space is allowed means that on smaller scales (like LVK), there could still be some quantum signatures left due to decoherence not completing before the end of inflation. 

The results presented here show the first rigorous observational constraints on quantum decoherence scenarios for tensor modes, and they highlight the importance of taking quantum decoherence effects into account when studying the origin of the observed GW background. 
In this work, we consider only the effect on the GW power spectrum from the quantum decoherence of tensor modes. However, future work will focus on observationally constraining the quantum decoherence of both scalar and tensor modes at the same time. This would increase the studied parameter space considerably, but it would shed light on the interplay of different quantum effects.

\appendix
\section{Varying \texorpdfstring{$\Lambda$}{TEXT}CDM parameters}\label{sec: LCDM}

\begin{figure}
    \centering
    \includegraphics[width=\linewidth]{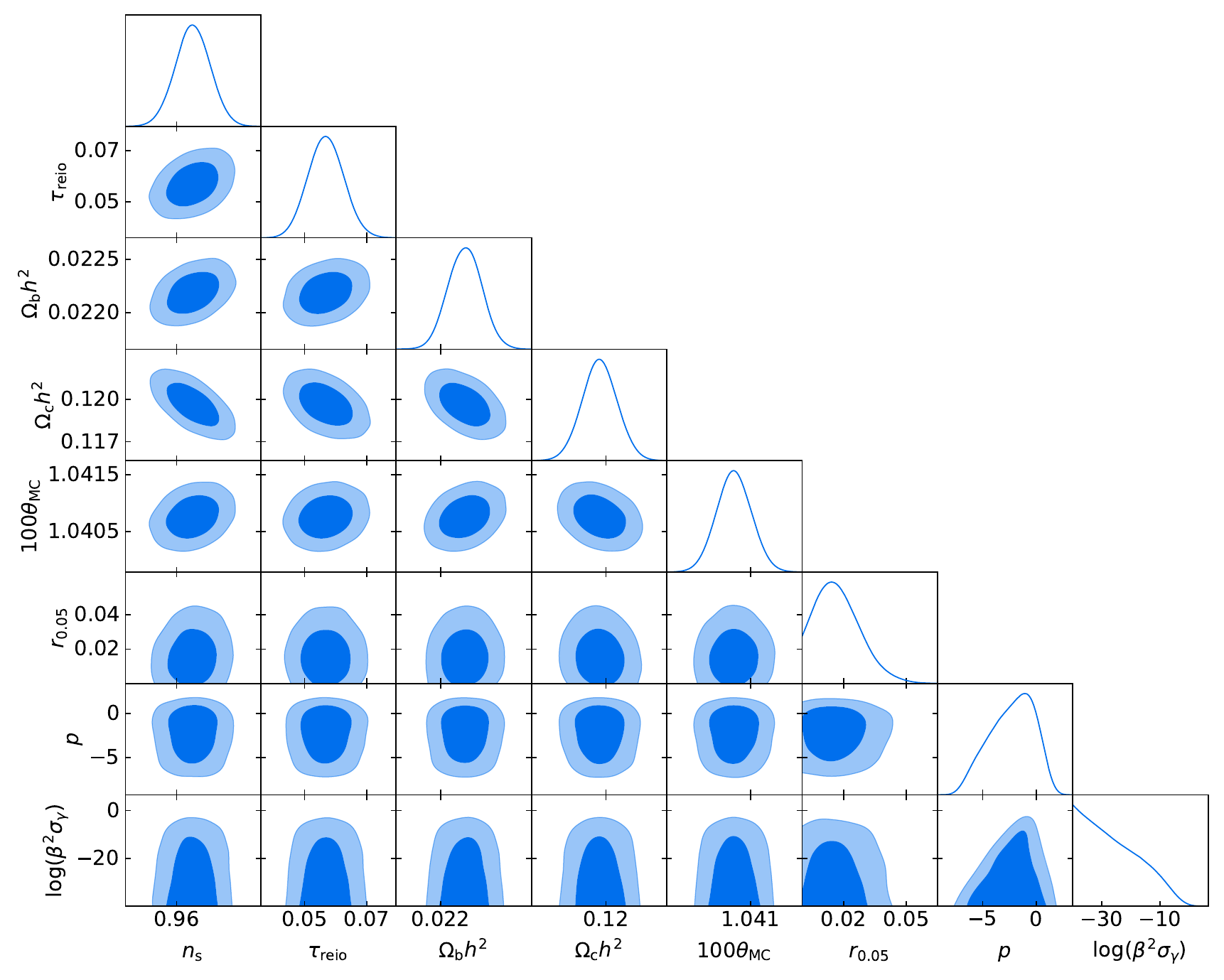}
    \caption{The 1D and 2D 68\% and 95\% CL interval constraints on $A_s, n_s, \tau_{\rm reio}, \Omega_bh^2, \Omega_ch^2, 100 \theta_{s}, r_{0.05}, p, \beta^2\sigma_\gamma$, using PL+BK18+LVK and the NPIPE high-$\ell$ CamSpec TTTEEE likelihood. }
    \label{fig: LCDM}
\end{figure}

Let us elaborate on the impact of varying the $\Lambda$CDM parameters on the results obtained in this work. To further constrain the scalar parameters, we impose the NPIPE high-$\ell$ CamSpec TTTEEE spectrum~\cite{Rosenberg_2022}, in addition to the constraints used above. The interval constraints are shown in Fig.~\ref{fig: LCDM}, and the bounds obtained on $r_{0.05}$ and the quantum decoherence parameters are
\begin{equation}
    r_{0.05} < 0.037, \ \log(\beta^2\sigma_\gamma) < -8.4, \ -6.3 < p < 1.2 \ \ (95\% \rm{CL}).
\end{equation}

It is clear from these bounds that varying the $\Lambda$CDM parameters does not significantly affect the results we obtain. The upper bound on $r_{0.05}$ is slightly larger than the above obtained results. Moreover, there is only a small difference in the bounds obtained for the decoherence parameters, probably due to the increase in parameter space. This small change does not affect the conclusions of this work in any way. Therefore, to avoid sampling a bigger parameter space, we do not vary the $\Lambda$CDM parameters in this work.

\section{Tilted standard power spectrum}\label{sec: r0.01}

\begin{figure}
    \centering
    \includegraphics[width=\linewidth]{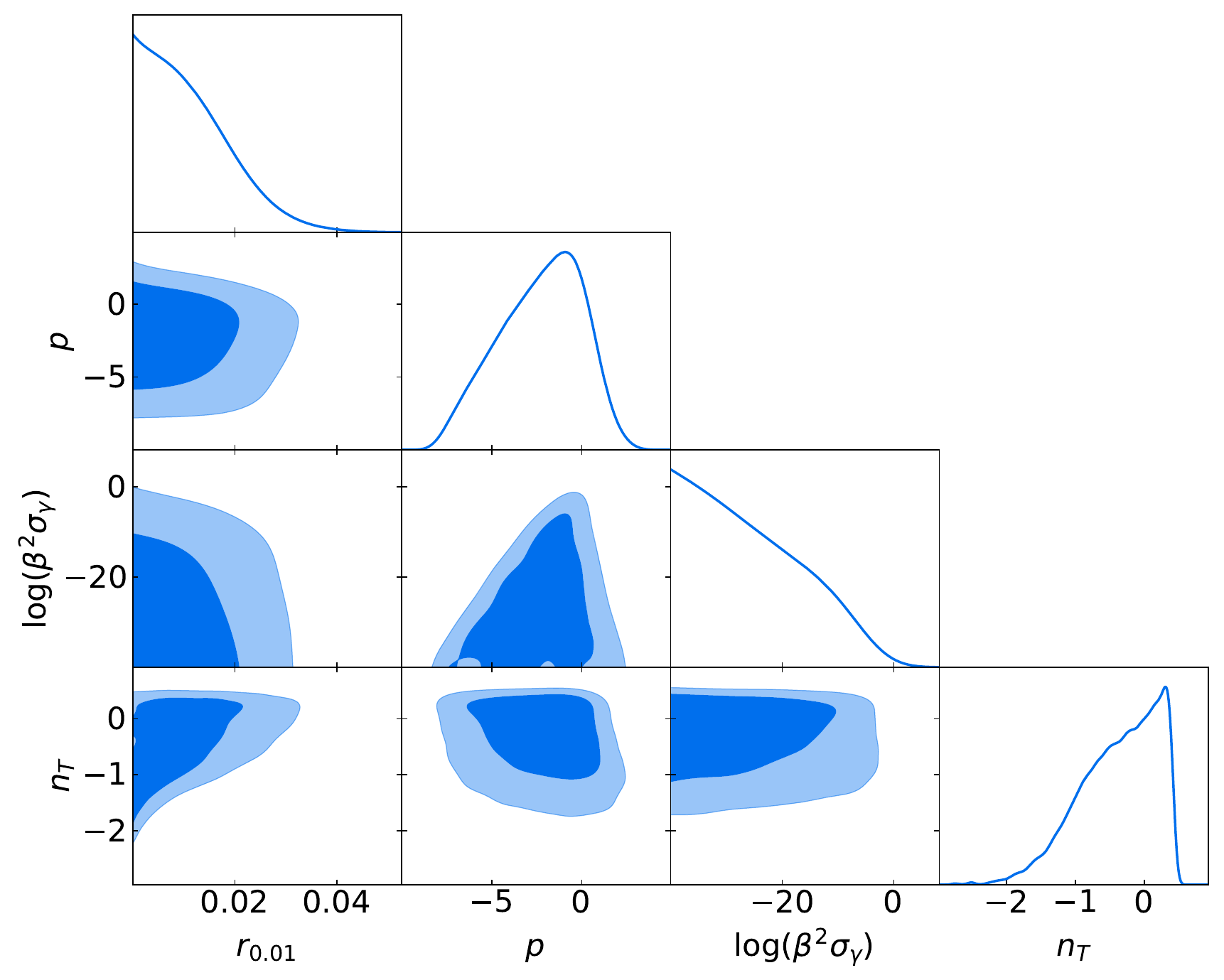}
    \caption{The 1D and 2D 68\% and 95\% CL interval constraints on $r_{0.01}, p, \beta^2\sigma_\gamma, n_T$, using PL+BK18+LVK. Additionally, we set $H_{\ast}l_E = 10^{-3}$, and $\Delta N_{\ast} = 50$.}
    \label{fig: tilted r_0.01}
\end{figure}

Throughout our entire work, we take the pivot scale to be $k_\ast = 0.05 \ \rm{Mpc}^{-1}$. However, to study if quantum decoherence has an effect on the marginalized constraints on the standard power spectra parameters $r$ and $n_T$ with respect to the literature \cite{Galloni_2023}, we change our pivot scale to $k_\ast = 0.01 \ \rm{Mpc}^{-1}$. This gives us the constraints
\begin{equation}
    r_{0.01} < 0.028, \  -1.44 < n_T < 0.46 \ \ (95\% \rm{CL}).
\end{equation}
Compared to the results obtained in~\cite{Galloni_2023}, both the bounds on $r_{0.01}$ and $n_T$ show no significant difference. This is consistent with the other results presented in this paper, where the upper bound on $r_{0.05}$ remains consistent with existing literature (see \ref{sec: results r, p, sigma}). Especially the 1D posteriors (as shown in Fig.~\ref{fig: tilted r_0.01}, and Fig. 12 of~\cite{Galloni_2023}), reflect exactly the same behavior for both results.

The bounds on the quantum decoherence parameter are
\begin{equation}
    \log(\beta^2\sigma_\gamma) < -7.2, \  -6.9 < p < 1.6 \ \ (95\% \rm{CL}).
\end{equation}
These bounds show a minor shift with respect to the results for $r_{0.05}$, as is to be expected. This highlights the impact of the pivot scale on the quantum decoherence power spectrum, which is small but not negligible.

\begin{acknowledgments}
We would like to thank Sabino Matarrese, Nicola Bellomo, and Vincent Vennin for the helpful discussions. Additionally, we would like to thank Jesus Torrado for his help with the \texttt{cobaya} code. 
This work is partially supported by the MUR Departments of Excellence grant “Quantum Frontiers” of the Physics and Astronomy Department of Padova University. NB and GG acknowledge financial support from the COSMOS network (www.cosmosnet.it) through the ASI (Italian Space Agency) Grants 2016-24-H.0, 2016-24-H.1-2018 and 2020-9-HH.0. GG and NB acknowledge the support by the MUR PRIN2022 Project “BROWSEPOL: Beyond standaRd mOdel With coSmic microwavE background POLarization”-2022EJNZ53 financed by the European Union - Next Generation EU.

\end{acknowledgments}

\appendix

\bibliographystyle{JHEP}
\bibliography{biblio.bib}

\end{document}